\documentclass{article}
\usepackage[utf8]{inputenc}
\usepackage{authblk}
\usepackage[dvips]{graphicx}
\graphicspath{{noiseimages/}}
\usepackage{xcolor}
\usepackage[T2A]{fontenc}
\usepackage{tikz}
\usepackage{pgfplots}
\usepackage[a4paper, left=2cm,right=2cm,top=2cm,bottom=2cm,bindingoffset=0cm]{geometry}
\pgfplotsset{compat=1.7}
\usetikzlibrary{intersections, pgfplots.fillbetween}
\usetikzlibrary{decorations.pathmorphing}
\usepackage[mode=buildnew]{standalone}
\usepackage[toc,page]{appendix}
\usepackage{amsmath}
\usepackage{amssymb}
\usepackage{float}
\usepackage{empheq}
\usepackage[framemethod=tikz]{mdframed}
\usepackage{mathtools}
\usepackage{comment}
\usepackage[english]{babel}
\usepackage{hyperref}
\definecolor{urlcolor}{HTML}{990000}
\definecolor{linkcolor}{HTML}{005F5F} 
\hypersetup{pdfstartview=FitH,  linkcolor=linkcolor,urlcolor=urlcolor, colorlinks=true,citecolor=blue}
\usepackage[page]{appendix}

\newcommand{\w}{\omega}

\renewcommand{\hat}{}

\newmdenv[innerlinewidth=0.5pt, roundcorner=4pt,linecolor=black,innerleftmargin=6pt,
innerrightmargin=6pt,innertopmargin=3pt,innerbottommargin=6pt]{mybox}

\title{Loop corrections to a cosmological particle creation}
\author[1,2]{E.T.Akhmedov}
\author[1,2]{P.A.Anempodistov\footnote{\tt anempodistov.pa@phystech.edu }}
\affil[1]{Institutskii per. 9, Moscow Institute of Physics and Technology, 141700, Dolgoprudny, Russia}
\affil[2]{B. Cheremushkinskaya, 25, Institute for Theoretical and Experimental Physics, 117218, Moscow, Russia}

\begin{document}

\maketitle

\begin{abstract}
    We consider dynamics of the massive minimally coupled scalar field theory in an expanding Friedmann–Lemaître–Robertson–Walker universe. We consider the standard toy model of the conformally flat space-time where the conformal factor becomes constant at the distant past and the distant future. Employing Schwinger-Keldysh diagrammatic technique, we compute infrared loop corrections to the occupation number and anomalous quantum average of the scalar field and show that these corrections are growing with time. Using these observations, we demonstrate that the regularized stress-energy tensor at the distant future acquires substantial quantum corrections which exceed the long known tree-level contributions to the particle flux. 
\end{abstract}

\newpage

\tableofcontents

\newpage 

\section{Introduction}

In this paper we consider massive minimally coupled scalar quantum field theory in an expanding Friedmann–Lemaître–Robertson–Walker universe. For simplicity we mostly consider two-dimensional situation, but also briefly discuss the situation in higher dimensions in the Appendix. (In higher dimensions the situation appears to be very similar.) The universe that we consider has the expanding geometry, which is sandwiched between flat space-time regions at past and future infinities. Namely the conformal factor of the metric continuously changes from one constant in the remote past to another constant in the remote future. Due to such a geometry of space-time there is a particle creation. Namely in the Gaussian approximation, i.e. for the fields without self-interactions, this model was extensively studied long ago and the particle creation was rate was calculated in \cite{Birrell:1982ix, Bernard:1977pq}. 

The aim of this paper is to investigate the behavior of self-interacting fields in this background. We consider massive scalar field theory with $\lambda \, \phi^4$ self-interaction. Based on the intuition gained by the previous work of our group we expect that loop corrections to the stress-energy fluxes will grow with time and overcome the tree-level contribution \cite{Akhmedov:2021rhq}. That is due to strong infrared (secular or memory) corrections, which appear even for massive fields due to non-stationary situation in the system under consideration, which is caused by the time-dependent metric. In fact, below we find that the stress-energy flux which was found in \cite{Birrell:1982ix, Bernard:1977pq} receives strong (growing with time or secular) loop corrections.

%For example, for a scalar field with $\lambda \phi^4$ self-interaction in the $(d+1)$-dimensional flat spacetime one obtains that the two-loop corrections to the propagators yield the following contribution to the occupation number \cite{Akhmedov:2013vka, Berges:2004yj}:
%\begin{multline} \label{n_p_flat}
 %       n_p^{(2)} \sim \lambda^2 (t-t_0) \int \frac{d^dq_1 d^dq_2 d^dq_3}{\w_{q_1}\w_{q_2}\w_{q_3} } \bigg[ (1+n_{p})(1+n_{q_1})n_{q_2}n_{q_3} - n_{p}n_{q_1}(1+n_{q_2})(1+n_{q_3}) \bigg] \times \\ \times \delta(\w_p+\w_{q_1}-\w_{q_2}-\w_{q_3}),
  %  \end{multline}
%where $\w_q = \sqrt{\vec{q}^2+m^2}$. Here the 'collision integral' in the r.h.s. the first term in the square brackets describes the process of scattering of two particles with initial momenta $\vec{p}$ and $\vec{q}_1$ and with final momenta $\vec{q}_2$ and $\vec{q}_3$, while the second term in the square brackets describes reverse process. 

Secularly growing loop corrections occur in various strong backgrounds and the character of the secular growth is sensitive to the choice of the background geometry and to the choice of the initial state and even the Cauchy surface. In fact, one encounters such a phenomenon in cosmological backgrounds  \cite{Akhmedov:2013vka}, \cite{Akhmedov:2019cfd} in the black hole collapse background \cite{Akhmedov:2015xwa}, in strong electric field backgrounds \cite{Akhmedov:2014doa}, \cite{Akhmedov:2014hfa}, in the backgrounds of moving mirrors \cite{Akhmedov:2017hbj}, \cite{Trunin:2021fom}, \cite{Akopyan:2020xqu} and even in scalar field backgrounds \cite{Akhmedov:2020haq}, \cite{Akhmedov:2019ubc}. There is an evidence that secular growth of loop corrections in non-stationary situations or in background fields is quite a generic phenomenon \cite{Akhmedov:2021rhq} (see, however, \cite{Lanina:2020yvh}, \cite{Akhmedov:2019rvx}, \cite{Akhmedov:2021agm}, \cite{Akhmedov:2017dih} for some specific situations). 

Concretely in this paper we calculate loop corrections to the two point correlation functions and extract from them corrections to the occupation (or level population) numbers $Tr[\overset{\wedge}{\rho}a^+_{\Vec{q}}a_{\Vec{q}'}]$ and to the anomalous quantum averages (or anomalous expectation values) $Tr[\overset{\wedge}{\rho}a_{\Vec{q}}a_{\Vec{q}'}]$. (Here $\overset{\wedge}{\rho}$ is a density matrix characterizing the state of the theory, $a^+_{\Vec{q}}$ and $a_{\Vec{q}}$ are creation and annihilation operators in the theory and the trace ``Tr'' is taken over the Fock space of the theory.) Then we investigate how the regularized stress-energy tensor changes due to the quantum loop generation of the level population and anomalous expectation value. We find that loop corrections to the latter quantities grow as the average time of the two-point functions is taken to the future infinity.

The structure of the paper is as follows: in Section \ref{setup} we describe the background geometry, define modes, and construct propagators for the Schwinger-Keldysh diagrammatic technique for generic spatially homogeneous states. In Section \ref{ir_loops} we compute infrared loop corrections to the occupation number and anomalous quantum average and demonstrate that $n_p$ and $\kappa_p$ acquire growing with time corrections. In the concluding Section we calculate the regularized stress-energy tensor and find contributions to it due to the self-interaction of the field. In the Appendix we discuss some issues concerning the generalization of our considerations to higher dimensions.

%\newpage

\section{Setup for the problem} \label{setup}

\subsection{Space-time geometry and modes}

We consider expanding universe with the following metric
\begin{equation}
    ds^2 = C(\eta) \, \Big(d \eta^2 -dx^2\Big),\label{metric1}
\end{equation}
where the conformal factor is given by \cite{Birrell:1982ix}, \cite{Bernard:1977pq}:
\begin{equation} \label{conformal_factor}
    C(\eta) = A + B \tanh(\rho \eta),
\end{equation}
with some positive constants $A > B$, and $\rho$. In the main text of this paper we consider the two-dimensional situation (to highlight its peculiarities) and then extend it to any dimensional case in the Appendix. The situation in higher dimensions is not very much different.

%\begin{figure}[H]
%\includegraphics[width=8cm]{conf_factor.pdf}
%\centering
%\caption{}
%\label{metric_plot}
%\end{figure}

We consider massive minimally coupled scalar field on this gravitational background with the action given by
\begin{equation} \label{bare_action}
    S = \int d^2 x \sqrt{-g} \, \Big[ \frac{1}{2} \, (\partial_{\mu} \phi )^2 - \frac{m^2}{2} \phi^2 - \frac{\lambda}{4!} \, \phi^4\Big].
\end{equation}
%where the $S_{int}$ contains quartic, $\phi^4$, self-interaction term.
As the background geometry is spatially homogeneous, it is convenient to separate the variables as:
\begin{equation} \label{separation_variables}
    u_k(\eta,x) = e^{ikx} g_k(\eta).
\end{equation}
Then, for a free field the equation for the temporal part of the mode function is
\begin{equation}
    \frac{d^2}{d\eta^2} g_k(\eta) + \bigg[k^2 + C(\eta) \, m^2 \bigg] g_k(\eta) =0.
\end{equation}
%whose generic solution is:
%\begin{multline}
%    g_k(\eta) = \Tilde{C}_1 \, e^{-i \w_+ \eta} e^{-\frac{i\w_-}{\rho} \log \cosh(\rho \eta) } F \bigg( 1+\frac{i\w_-}{\rho} ,  \frac{i\w_-}{\rho} ; 1- \frac{i \w_{in}}{\rho} ; \frac{1+\tanh (\rho \eta)}{2} \bigg)+\\+\Tilde{C}_2 \, e^{-i \w_- \eta} e^{-\frac{i\w_+}{\rho} \log \cosh(\rho \eta) } F \bigg( 1+\frac{i\w_+}{\rho} ,  \frac{i\w_+}{\rho} ; 1+ \frac{i \w_{in}}{\rho} ; \frac{1+\tanh (\rho \eta)}{2} \bigg).
%\end{multline}
Among all possible solutions of the equation above one distinguishes the in-modes that are given by
\begin{equation} \label{in_modes}
    u_k^{in}(\eta,x) = (4\pi \w_{in})^{-\frac{1}{2}} e^{ikx-i\w_{+}\eta -i\frac{\w_-}{\rho} \log[2\cosh(\rho \eta)] } F\bigg(1+\frac{i\w_-}{\rho},\frac{i\w_-}{\rho};1-\frac{i\w_{in}}{\rho};\frac{1+\tanh(\rho \eta)}{2}\bigg),
\end{equation}
where $F(a,b\,;c\,;z)$ is the Gaussian hypergeometric function and
\begin{align}
    \w_{in}(k) = \sqrt{k^2+m^2 \, (A-B)}, \quad
    \w_{out}(k) = \sqrt{k^2+m^2 \, (A+B)}, \quad {\rm and} \quad
    \w_{\pm} =\frac{1}{2} (\w_{out} \pm \w_{in}).
\end{align}
%In fact, using that
%\begin{multline}
 %   F\bigg(1+\frac{i\w_-}{\rho},\frac{i\w_-}{\rho};1-\frac{i\w_{in}}{\rho};\frac{1+\tanh(\rho \eta)}{2}\bigg) = \frac{\Gamma(1-i \w_{in}/\rho) \Gamma(-i\w_{out}/\rho)  }{ \Gamma(-i\w_+ / \rho) \Gamma(1-i\w_+/\rho) } \\ F\bigg(1+\frac{i\w_-}{\rho},\frac{i\w_-}{\rho};1+\frac{i\w_{out}}{\rho};\frac{1-\tanh(\rho \eta)}{2}\bigg) + \frac{ \Gamma(1-i\w_{in}/\rho) \Gamma(i\w_{out}/\rho) }{ \Gamma(1+i\w_- / \rho) \Gamma(iw_- / \rho) } \bigg( \frac{1-\tanh(\rho \eta)}{2} \bigg)^{-\frac{i\w_{out}}{\rho}} \\ F\bigg(1-\frac{i\w_+}{\rho},-\frac{i\w_+}{\rho};1-\frac{i\w_{out}}{\rho};\frac{1-\tanh(\rho \eta)}{2}\bigg),
%\end{multline}
Using tranformation formulas for the hypergeometric function, one can show that the in-modes \eqref{in_modes} have the following asymptotics
\begin{align} \label{in_modes_asympt}
    u_k^{in}(\eta,x) = e^{ikx} g_k^{in}(\eta) \approx
    \begin{cases}
        \frac{1}{\sqrt{4\pi \w_{in}}} e^{ikx-i\w_{in}\eta}, \qquad  &\text{as} \quad \eta \to -\infty \\
        \frac{1}{\sqrt{4\pi \w_{in}}} e^{ikx} \bigg[ \alpha(k) \,  e^{-i\w_{out}\eta} + \beta(k) \, e^{i\w_{out}\eta} \bigg], \qquad &\text{as} \quad \eta \to +\infty,
    \end{cases}
\end{align}
where 
\begin{align} \label{c_1_c_2}
    \alpha(k) = \frac{\Gamma(1-i \w_{in}/\rho) \Gamma(-i\w_{out}/\rho)  }{ \Gamma(-i\w_+ / \rho) \Gamma(1-i\w_+/\rho) }, \qquad
    \beta(k) = \frac{ \Gamma(1-i\w_{in}/\rho) \Gamma(i\w_{out}/\rho) }{ \Gamma(1+i\w_- / \rho) \Gamma(i\w_- / \rho) }.
\end{align}
From \eqref{in_modes_asympt} one can see that in the flat asymptotic past ($\eta \to -\infty$) the in-modes correspond to the usual notion of the particle in flat space-time, because they are described by single waves. That is the reason why they are referred to as in-modes.

The mode decomposition of the field operator is
\begin{equation}
    \phi(\eta,x) = \int^{\infty}_{-\infty} dk \Big[ a_k u_k^{in}(\eta,x) + a_k^{\dagger} u_k^{in*}(\eta,x) \Big].
\end{equation}
Using the properties of the hypergeometric functions, one can show that the canonical commutation relations for the field $\phi$ with its conjugate momentum and for $a_k$ with $a_k^{\dagger}$ are satisfied.

Similarly, one can define out-modes as
\begin{equation} \label{out_modes}
    u_k^{out}(\eta,x) = (4\pi \w_{in})^{-\frac{1}{2}} e^{ikx-i\w_{+}\eta -i\frac{\w_-}{\rho} \log[2\cosh(\rho \eta)] } F\bigg(1+\frac{i\w_-}{\rho},\frac{i\w_-}{\rho};1+\frac{i\w_{out}}{\rho};\frac{1-\tanh(\rho \eta)}{2}\bigg).
\end{equation}
They correspond to the usual notion of the particle in flat space-time in the flat asymptotic future ($\eta \to +\infty$). In fact, they have the following asymptotic behavior:
\begin{align} \label{out_modes_asympt}
    u_k^{out}(\eta,x) = e^{ikx} \, g_k^{out}(\eta) \approx
    \begin{cases}
        \frac{1}{\sqrt{4\pi \w_{out}}} e^{ikx} \bigg[ \gamma(k) \,  e^{-i\w_{in}\eta} + \delta(k) \, e^{i\w_{in}\eta} \bigg], \qquad &\text{as} \quad \eta \to -\infty \\
        \frac{1}{\sqrt{4\pi \w_{out}}} e^{ikx-i \w_{out}\eta} , \qquad &\text{as} \quad \eta \to +\infty,
    \end{cases}
\end{align}
where
\begin{align}
    \gamma(k)=\frac{\Gamma(1+i \w_{out}/\rho) \Gamma(i\w_{in}/\rho)  }{ \Gamma(i\w_+ / \rho) \Gamma(1+i\w_+/\rho) }, \qquad
    \delta(k)=\frac{ \Gamma(1+i\w_{out}/\rho) \Gamma(-i\w_{in}/\rho) }{ \Gamma(1+i\w_- / \rho) \Gamma(i\w_- / \rho) }.
\end{align}
The mode decomposition of the field operator in these modes we will denote as follows:
\begin{equation}
    \phi(\eta,x) = \int^{\infty}_{-\infty} dk \Big[ b_k u_k^{out}(\eta,x) + b_k^{\dagger} u_k^{out*}(\eta,x) \Big].
\end{equation}
We find it instructive to calculate loop corrections both for in- and out- Fock space states separately.

\subsection{Schwinger-Keldysh formalism}

Below we will show that loop corrections to the expectation value of the stress-energy tensor grow with time, thus, signaling substantial amplification of the particle creation rate by quantum loop corrections. This is the generic situation in time-dependent or strong background fields, as was explained in the Introduction.
The background that we consider here (\ref{metric1}), (\ref{conformal_factor}) is time-dependent. Hence, the free Hamiltonian of the scalar theory under consideration (\ref{bare_action}) also depends on time. Hence, we are dealing here with an example of non-stationary situation. In such a case to calculate loop corrections one has to apply the Schwinger-Keldysh rather than Feynman diagrammatic technique (see e.g. \cite{LL10} \cite{Berges:2004yj}, \cite{Kamenev} for the introduction into the subject).

\begin{figure}[h]
\includegraphics[width=16cm]{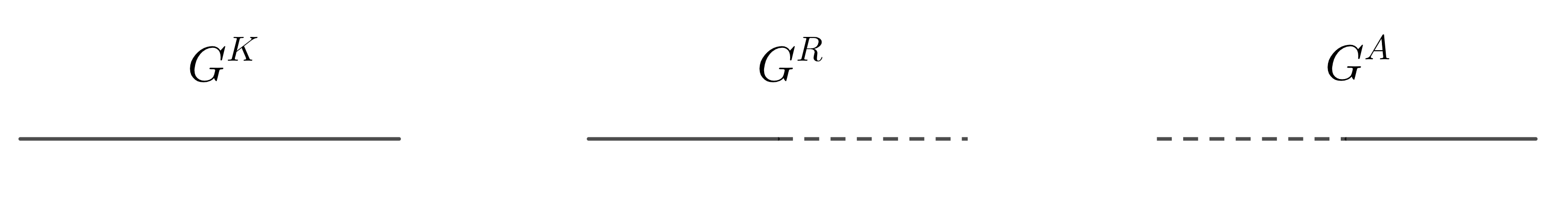}
\centering
\caption{Graphical notation for the Keldysh, retarded, and advanced propagators respectively.}
\label{tree_diagr}
\end{figure}

In the Schwinger-Keldysh formalism the system is considered to evolve on a closed time contour which consists of two branches: forward branch (from $-\infty$ to $+\infty$) and backward branch (from $+\infty$ to $-\infty$). We denote the field operator on the forward and backward branches of the contour as $\phi^+$ and $\phi^-$, correspondingly. Then, one can construct four types of propagators, but only three of them are linearly independent. To transform to these three linearly independent propagators, one performs the Keldysh rotation:
\begin{align}
    &\phi^{cl}=\frac{1}{2} (\phi^++\phi^-), \nonumber \\
    &\phi^q = \phi^+-\phi^-.
\end{align}
Then, the three propagators are the Keldysh, retarded, and advanced two-point functions correspondingly:
\begin{align}
    %&G^>(x_1,x_2) = \langle \phi(x_1) \phi(x_2) \rangle, \qquad G^<(x_1,x_2) = \langle \phi(x_2) \phi(x_1) \rangle, \nonumber \\
    &G^K(\eta,x|\eta',x') \equiv \Big\langle \phi^{cl}(\eta,x)\, \phi^{cl}(\eta',x') \Big\rangle = \frac{1}{2} \Big\langle \Big\{\phi(\eta,x), \, \phi(\eta',x')\Big\} \Big\rangle, \nonumber\\
    &G^R(\eta,x|\eta',x') \equiv \Big\langle \phi^{cl}(\eta,x) \, \phi^{q}(\eta',x') \Big\rangle = \theta(\eta-\eta')\,  \Big[ \phi(\eta,x), \, \phi(\eta',x')\Big] , \nonumber\\
    &G^A(\eta,x|\eta',x') \equiv \Big\langle \phi^{q}(\eta,x) \, \phi^{cl}(\eta',x') \Big\rangle = -\theta(\eta'-\eta)\,  \Big[ \phi(\eta,x), \, \phi(\eta',x') \Big] . \label{KRA}
\end{align}
Graphically, these propagators are denoted as shown on the Fig. \ref{tree_diagr}

For a generic state the following mode expansion is valid %for 
%the Wightman function:
%\begin{multline}
 %   G^>(\eta,x|\eta',x') \equiv \Big\langle \phi(\eta, x) \phi(\eta',x') \Big\rangle = \int dkdq \bigg[ \Big(\delta_{kq} + \langle a_q^{\dagger} a_k \rangle\Big) \, u_k^{in}(\eta,x) \, u_q^{in*}(\eta',x') +\\+ \langle a^{\dagger}_k a_q \rangle \, u^{in*}_k(\eta,x) \, u_q^{in} (\eta',x') + \langle a_k a_q \rangle \, u_k^{in} (\eta,x) \,  u_q^{in} (\eta',x') + \langle a^{\dagger}_k a^{\dagger}_q \rangle \, u^{in*}_k(\eta,x) \,  u^{in*}_q(\eta',x') \bigg],
%\end{multline}
for the Keldysh propagator:
\begin{multline} \label{keldysh_generic}
    G^K(\eta,x|\eta',x') = \iint dk dq \, \bigg\{ \bigg( \frac{1}{2} \delta_{kq} + \Big\langle a_q^{\dagger} a_k \Big\rangle \bigg) \Big[ u_k^{in}(\eta,x) \, u_q^{in*}(\eta',x') + u_k^{in}(\eta',x') \, u_q^{in*}(\eta,x) \Big]+ \\+
     \Big\langle a_k a_q \Big\rangle \, u_k^{in} (\eta,x) \, u_q^{in} (\eta',x') + \Big\langle a^{\dagger}_k a^{\dagger}_q \Big\rangle \, u^{in*}_k(\eta,x) \, u^{in*}_q(\eta',x') \bigg\},
\end{multline}
and for the commutator:
\begin{equation} \label{g_modes}
    G(\eta,x|\eta',x') \equiv  \Big[ \phi(x), \, \phi(x') \Big]  = \int dk \Big[ u_k^{in}(\eta,x)\, u_k^{in*}(\eta',x') - u_k^{in}(\eta',x') \, u_k^{in*}(\eta,x) \Big],
\end{equation}
via which one can express the retarded and advanced propagators, as in (\ref{KRA}). Note that while the tree-level retarded and advanced propagators (and the field commutator) are state independent, the Keldysh propagator does depend on the state of theory.

Because the metric under consideration is spatially homogeneous we consider only spatially homogeneous initial states in which the following relations necessarily hold:
\begin{equation}
    \langle a^{\dagger}_q a_k \rangle \equiv {\rm Tr} \left(\hat{\rho} \, a^+_k \, a_q \right) = n_k\, \delta(k-q), \qquad \langle a_q a_k \rangle = \kappa_k\, \delta(k+q), \qquad \langle a^{\dagger}_q a^{\dagger}_k \rangle=\kappa_k^* \,\delta(k+q),
\end{equation}
where $n_k$ is the occupation (or level population) number and $\kappa_k$ is the anomalous quantum expectation value (or anomalous average). For such states the Keldysh propagator, which is sensitive to the state of the theory, simplifies to
\begin{multline} \label{gk_modes}
    G^K(\eta,x|\eta',x') = \int^{\infty}_{-\infty} dk \, \bigg\{ \bigg( \frac{1}{2} + n_k \bigg) \Big[ u_k^{in}(\eta,x) \, u_k^{in*}(\eta',x') + u_k^{in}(\eta',x') \, u_k^{in*}(\eta,x) \Big] + \\ + \kappa_k \, u_k^{in} (\eta,x) \, u_{-k}^{in} (\eta',x') + \kappa_k^* \, u_k^{in*} (\eta,x) \, u_{-k}^{in*} (\eta',x') \bigg\}.
\end{multline}
Using \eqref{separation_variables}, one can define the Fourier components of the propagators as
\begin{align}
    &G^K(\eta,x|\eta',x') = \int^{\infty}_{-\infty}  dk \, e^{ik(x-x')} \, G^K(k|\eta,\eta').
    %&G^K(k|\eta,\eta') = \int^{\infty}_{-\infty} \frac{d(x-x')}{2\pi} e^{-ik(x-x')} G^K(\eta,x|\eta',x').
\end{align}
The Fourier components of \eqref{gk_modes} and \eqref{g_modes} are explicitly given by
\begin{gather} 
    G^K(k|\eta,\eta') = \bigg( \frac{1}{2}+n_k \bigg) \bigg( g_k^{in}(\eta)g_k^{in*}(\eta')+  g_k^{in*}(\eta)g_k^{in}(\eta') \bigg) + \kappa_k \,  g_k^{in} (\eta) g_{k}^{in} (\eta') + \kappa_k^* \, g_k^{in*} (\eta) g_{k}^{in*} (\eta') , \nonumber\\
    \text{and} \qquad  G(k|\eta,\eta') =  g_k^{in}(\eta)g_k^{in*}(\eta')- g_k^{in*}(\eta)g_k^{in}(\eta'), \label{gk_modes_in}
\end{gather}
where we have restricted ourselves to the case in which occupation number is symmetric under the inversion of the spatial momentum: $n_{k} = n_{|k|}.$

Alternatively, one can quantize the field using out-modes \eqref{out_modes}, i.e. decompose the field operator as
\begin{equation}
    \phi(\eta,\Vec{x}) = \int^{\infty}_{-\infty} dk \Big[ b_k u_k^{out}(\eta,\vec{x}) + b_k^{\dagger} u_k^{out*}(\eta,\vec{x}) \Big],
\end{equation}
and consider spatially homogeneous initial states in which
\begin{equation}
    \langle b^{\dagger}_q b_k \rangle = \bar{n}_k\, \delta(k-q), \qquad \langle b_q b_k \rangle = \bar{\kappa}_k\, \delta(k+q), \qquad \langle b^{\dagger}_q b^{\dagger}_k \rangle=\bar{\kappa}_k^* \,\delta(k+q).
\end{equation}
In such a case, the Keldysh propagator has the same form as (\ref{gk_modes_in}), but with $g_k^{in}(\eta)$, $n_k$, $\kappa_k$ replaced by $g_k^{out}(\eta)$, $\bar{n}_k$, and $\bar{\kappa}_k$ correspondingly.
%\begin{multline} \label{gk_modes_out}
 %   G^K(k|\eta,\eta') = \bigg( \frac{1}{2}+\Tilde{n}_k \bigg) \bigg( g_k^{out}(\eta)g_k^{out*}(\eta')+  g_k^{out*}(\eta)g_k^{out}(\eta') \bigg) +\\+ \Tilde{\kappa}_k g_k^{out} (\eta) g_{k}^{out} (\eta') + \Tilde{\kappa}_k^* g_k^{out*} (\eta) g_{k}^{out*} (\eta'),
%\end{multline}
At the same time the commutator function is equal to
\begin{equation} \label{g_modes_out}
    G(k|\eta,\eta') =  g_k^{out}(\eta)g_k^{out*}(\eta')- g_k^{out*}(\eta)g_k^{out}(\eta').
\end{equation}
One can show that this commutator function is the same as in (\ref{gk_modes_in}), because the transformation from the basis of in-modes to the out-modes is the canonical (simplectic) one.

Of course one can consider more generic states, which are belonging to Fock spaces constructed with the uses of modes that are related to the in- and out-harmonics via generic canonical transformations. One just has to pay attention to the Hadamard properties of the propagators, because singularities of the propagators are measurable, e.g., in the running of coupling constants. To keep task small in this paper we restrict our considerations only to the in- and out- Fock spaces.

\section{Infrared loop corrections} \label{ir_loops}

In this section we compute loop corrections to the occupation number and anomalous quantum expectation value in self-interacting scalar field theory. We consider quartic self-interaction, ${\lambda} \phi^4$, and show that for the in-modes the occupation number and anomalous quantum average are secularly growing with the average time of $\eta_1$ and $\eta_2$ of the Keldysh propagator $G^K(\eta_1,x_1|\eta_2,x_2)$. We also show that for the out-modes there are no growing with average time contributions to occupation number and anomalous quantum average. But we will see that for the out-modes there are secular divergences, i.e. for the out-state one cannot take the initial Cauchy surface to past infinity: otherwise loop correction will be infinite even after the implementation of the UV cutoff \cite{Akhmedov:2019cfd}.

\begin{figure}[H]
\includegraphics[width=10cm]{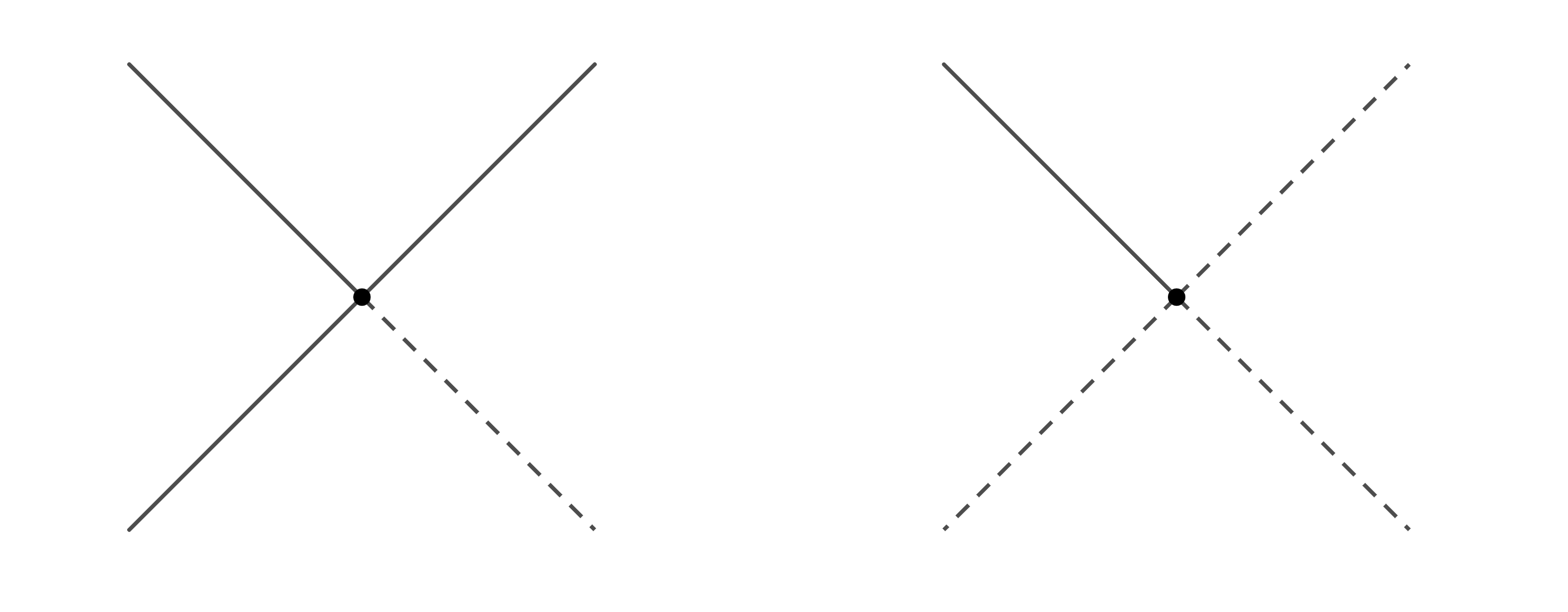}
\centering
\caption{Vertices for the quartic self-interaction.}
\label{vertices4}
\end{figure}

Consider now the theory with $\lambda \phi^4$ self-interaction:
\begin{equation}
    S_{int} = -\frac{\lambda}{4!} \int d^2x \sqrt{-g} \bigg( \phi_+^4 - \phi_-^4 \bigg) = -\frac{\lambda}{4!} \int d^2x \sqrt{-g} \bigg( 4 (\phi_{cl})^3 \phi_q + \phi_{cl} (\phi_q)^3 \bigg).
\end{equation}
One can see that there are two types of vertices in this case, which are depicted on the Fig. \ref{vertices4}. In the next two subsections we will separately consider contributions of diagrams of two types. We are mainly interested in the loop corrections to the Keldysh propagator, because it is this two-point correlation function, which is sensitive to the evolution of the state of the theory\footnote{Relevant loop corrections to the retarded and advanced propagators, unlike those to the Keldysh propagator, do not grow as both of their arguments are taken to the future infinity. We discuss this point in grater details in the section on the sunset diagrams.} \cite{Kamenev}, \cite{Akhmedov:2021rhq}.

\subsection{Tadpole diagrams}

We start with the consideration of the one and two loop tadpole diagrams. Then we resum the leading tadpole contributions from all loops.

\begin{figure}[H]
\includegraphics[width=15cm]{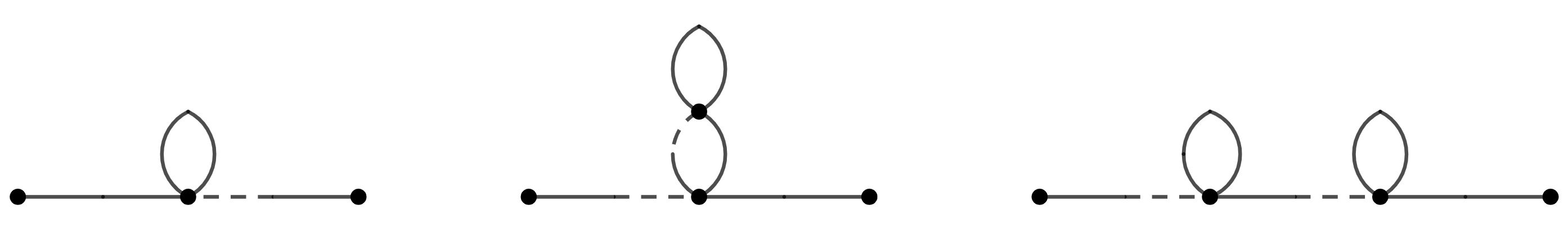}
\centering
\caption{A few leading tadpole diagrams. There are also diagrams of the same type with different placings of retarded and advanced propagators.}
\label{tadpole4}
\end{figure}

Some of the tadpole diagrams that contribute to the loop corrections of the Keldysh propagator are depicted on the Fig.\ref{tadpole4}. The sum of tadpole diagrams up to the two-loop level yields the following contribution:
\begin{multline} \label{G_k_tadpole}
    G^K_{\text{(2-tadpole)}}(\eta_1, x_1| \eta_2,x_2) = G^K_{12}-\frac{i\lambda}{2} \int^{\infty}_{\eta_0} d \eta_3\, dx_3\, C(\eta_3) \, \bigg( G^K_{13} G^K_{33} G^A_{32} + G^R_{13} G^K_{33} G^K_{32} \bigg) -\\- \frac{\lambda^2}{4} \int^{\infty}_{\eta_0} d \eta_3\, dx_3\, C(\eta_3) \int^{\infty}_{\eta_0} d \eta_4\, dx_4\, C(\eta_4) \, \bigg( 2 G^K_{13} G^R_{34} G^K_{34} G^K_{44} G^A_{32} + 2 G^R_{13} G^R_{34} G^K_{34} G^K_{44} G^K_{32} +\\+ G^K_{13} G^K_{33} G^A_{34} G^{K}_{44} G^A_{42} + G^R_{13} G^K_{33} G^K_{34} G^K_{44} G^A_{42} + G^R_{13} G^K_{33} G^R_{34} G^K_{44} G^K_{42} \bigg),
\end{multline}
where we have introduced notations of the form $G^K_{12} \equiv G^K(\eta_1,x_1|\eta_2,x_2)$ for the tree-level Keldysh propagator and similarly for the tree-level retarded and advanced propagators; $\eta_0$ here is the position of the initial Cauchy surface --- the time after which the self-interaction is adiabatically turned on. 

\textcolor{black}{At the same time the one-loop contributions to the retarded and advanced propagators are
%\begin{multline}
%    G^R_{(1)}(\eta_1, x_1| \eta_2,x_2) = - \frac{i\lambda}{2} \int^{\infty}_{\eta_0} d \eta_3\, dx_3\, C(\eta_3) \,G^R_{13} G^K_{33} G^R_{32}= \\=- \frac{i\lambda}{2} \int^{\infty}_{\eta_0} d \eta_3 \,dx_3\, C(\eta_3)\, \theta(\eta_1-\eta_3) \theta(\eta_3-\eta_2) G_{13} G^K_{33} G_{32}.
%\end{multline}
\begin{align} \label{G_R_A_tadpole}
    G^R_{(0+1)}(\eta_1, x_1| \eta_2,x_2) =G^R_{12} - \frac{i\lambda}{2} \int^{\infty}_{\eta_0} d \eta_3\, dx_3\, C(\eta_3) \,G^R_{13} G^K_{33} G^R_{32}, \nonumber \\
    G^A_{(0+1)}(\eta_1, x_1| \eta_2,x_2) =G^A_{12} - \frac{i\lambda}{2} \int^{\infty}_{\eta_0} d \eta_3\, dx_3\, C(\eta_3) \,G^A_{13} G^K_{33} G^A_{32}.
\end{align}
%In the limit \eqref{limit} (i.e $\eta_1 \approx \eta_2 \to \infty$), the one-loop contribution has a multiplier $\theta(\eta-\eta_3) \theta(\eta_3-\eta)=0$, so that at the one-loop there is no secular growth in the IR limit.
%\textcolor{red}{Я не понял почему надо смотреть поправки в пропагатор Келдыша в четвертом порядке, а в запаздывающий и опережающий пропагаторы во втором?}
Here we restrict our attention to the second order in $\lambda$ corrections to the Keldysh propagator and to the first order in $\lambda$ to the retarded and advanced propagators. That is necessary to close the Dyson-Schwinger equation for the resumed Keldysh propagator. } 
%Combining \eqref{G_k_tadpole} with \eqref{G_R_A_tadpole} and using that
%\begin{equation}
%    2 G^R_{34} G^K_{34} G^K_{44} = G^R_{34} G^K_{34} G^K_{44} + G^K_{34} G^K_{44} G^A_{43},
%\end{equation}

Let us investigate how the tadpole diagrams depend on the time coordinates of the endpoints. We would like to single out the leading contributions in a certain limit. In particular, we are interested in the limit when both coordinates of the two-point function are taken to the future infinity:
\begin{align} \label{limit}
    \frac{\eta_1+\eta_2}{2} \equiv \eta \gg \eta_1 - \eta_2.
\end{align}
The consideration of this limit allows one to trace the destiny of the state of the theory as the time goes by.

Let us start from the diagrams of the first type depicted on the Fig. \ref{tadpole4}. Denoting their contribution to the Keldysh propagator as $G^K_{a}$, one finds
\begin{equation}
    G^K_{a}(\eta_1, x_1| \eta_2,x_2) = \frac{i\lambda}{2} \int^{\infty}_{\eta_0} d \eta_3\, \int^{+\infty}_{-\infty} dx_3\, C(\eta_3) \, \bigg( G^K_{13} G^K_{33} G^A_{32} + G^R_{13} G^K_{33} G^K_{32} \bigg).
\end{equation}
Plugging here the expressions for the propagators \eqref{g_modes} and \eqref{gk_modes}, in the limit \eqref{limit} after the integration over $x_3$ one obtains\footnote{Here we have denoted the Fourier transformed Keldysh propagator $G^K(q| \eta_3,\eta_4)$ as $G^K_{q34}$.}:
\begin{multline} \label{type1_int}
    G^K_{a}(\eta_1, x_1| \eta_2,x_2) \approx -2 \pi i \lambda  \int^{\eta}_{\eta_0} d\eta_3 C(\eta_3) \int \int dp\, dr\, e^{ip(x_1-x_2)} G^K_{r33} \bigg( \frac{1}{2} + n_p \bigg) \times \\ \times \bigg\{ g_p^{in}(\eta_1) g_p^{in*}(\eta_3) g_p^{in*}(\eta_3) g_p^{in}(\eta_2) - g_p^{in*}(\eta_1) g_p^{in}(\eta_3) g_p^{in}(\eta_3) g_p^{in*}(\eta_2)  \bigg\}.
\end{multline}

\begin{figure}[H]
\includegraphics[width=8cm]{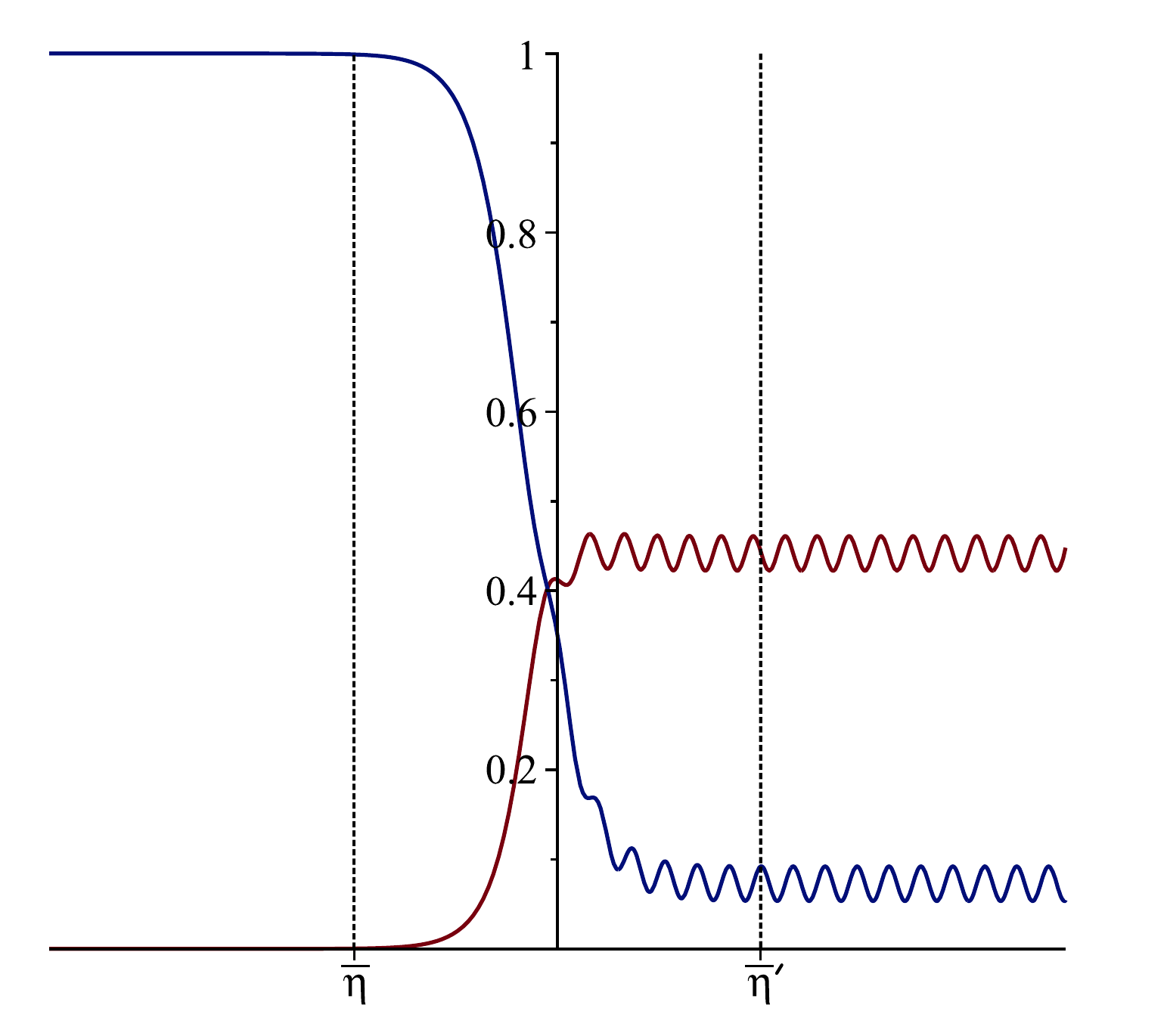}
\centering
\caption{Plot of the hypergeometric function $F(1-2i,-2i;1+i,{(1+\tanh(x))}/{2})$ against $x$: the blue line depicts the real part of the function, while the red line depicts the imaginary part. Vertical dotted lines characterize approximate positions of $\Bar{\eta}$ and $\Bar{\eta}'$.}
\label{mode_plot}
\end{figure}

Let us consider the integrals over $\eta_3$ and $\eta_4$ in \eqref{type1_int}. For now scientists do not known how to calculate such integrals exactly. To evaluate them approximately we divide the integration domain into three regions: remote past $(\eta_0,\Bar{\eta})$, intermediate expansion $(\Bar{\eta},\Bar{\eta}')$, and the remote future $(\Bar{\eta}',\eta)$. This division into three regions is done so that we could use asymptotic behavior \eqref{in_modes_asympt} of the in-modes in the past and future infinities (see Fig. \ref{mode_plot}). From the latter regions of space-time grwoing contributions can potentially come.

We can omit contributions from the remote past and intermediate regions, if the contribution from the future region grows with time in the limit \eqref{limit}, while the contribution from the remote past region does not grow as $\eta_0 \to - \infty$. This omission is justified if one keeps only the leading contribution to the integral in the limit \eqref{limit}, while the contribution from the remote past and intermediate regions remain finite. It is not very hard to check that the latter is true for the initial in-state. %(note that here we do not send $\eta_0$ to $-\infty$). 

To evaluate the contribution from the future region it is convenient to make the following change of variables in the double integral
\begin{equation}
    \int^{\eta}_{\Bar{\eta}'} d\eta_3 \int^{\eta}_{\Bar{\eta}'} d\eta_4 \, ... = \int^{\eta}_{\Bar{\eta}'} dt \int^{\eta - \Bar{\eta}'}_{\Bar{\eta}'-\eta} d \tau\, ... \, ,
\end{equation}
where we have introduced the new variables
\begin{equation}
    t = \frac{\eta_3+\eta_4}{2},\qquad  \tau = \eta_3-\eta_4.
\end{equation}
Then, in the limit \eqref{limit} one can approximate one of the integrals in \eqref{type1_int} as
\begin{multline} \label{integral}
    \int^{\eta}_{\eta_0} d\eta_3 \, C(\eta_3) \, G^K_{r33} \, g_p^{in*}(\eta_3) g_p^{in*}(\eta_3) \approx \int^{\eta}_{\Tilde{\eta}'} d\eta_3 \, C(\eta_3)\,  G^K_{r33} \,  g_p^{in*}(\eta_3) g_p^{in*}(\eta_3) \approx \\ \approx  \frac{(A+B)\, \eta}{4\pi^2 \w_{in}(r) \w_{in}(p)} \bigg( \frac{1}{2} + n_r \bigg)  \alpha^*(p) \beta^*(p) \bigg( |\alpha(r)|^2 + |\beta(r)|^2 \bigg), 
\end{multline}
where in the last step we have plugged the asymptotics of the in-modes at the future infinity \eqref{in_modes_asympt} and neglected $\Tilde{\eta}'$ in comparison with $\eta \to \infty$. In all, as one can see,
\begin{equation} \label{type1}
    G^K_{a} (\eta_1, x_1| \eta_2,x_2) \sim \lambda \, \eta, \quad {\rm as} \quad \eta \to +\infty,
\end{equation}
i.e. the contribution under consideration does grow with time and it comes from the future region of integration over $\eta_{3,4}$. 

We continue with the consideration of the diagrams of the second type depicted in Fig.\ref{tadpole4}. Denoting their contribution as $G^K_{b}$, one finds that
\begin{multline}
    G^K_{b} (\eta_1, x_1| \eta_2,x_2) = - \frac{\lambda^2}{2} \int^{\infty}_{\eta_0} d \eta_3\, dx_3\, C(\eta_3) \int^{\infty}_{\eta_0} d \eta_4\, dx_4\, C(\eta_4) \, \bigg(  G^K_{13} G^R_{34} G^K_{34} G^K_{44} G^A_{32} +  G^R_{13} G^R_{34} G^K_{34} G^K_{44} G^K_{32} \bigg) \approx \\ \approx - 8\pi^2 \lambda^2 \int^{\eta}_{\eta_0} d \eta_3\, C(\eta_3) \int^{\eta_3}_{\eta_0} d \eta_4\, C(\eta_4) \int dp \, dq \, ds \, e^{ip(x_1-x_2)} \bigg( \frac{1}{2}+n_p \bigg) \bigg( \frac{1}{2}+n_q \bigg) \bigg( \frac{1}{2}+n_s \bigg) \times \\ \times \bigg( g_p(\eta_1) g_p^*(\eta_3)  g_p^*(\eta_3) g_p(\eta_2) - g_p^*(\eta_1) g_p(\eta_3)  g_p(\eta_3) g_p^*(\eta_2) \bigg) \bigg( g_q(\eta_3)^2 g_q^*(\eta_4)^2 -  g_q^*(\eta_3)^2 g_q(\eta_4)^2 \bigg) |g_s(\eta_4)|^2.
\end{multline}
Calculating the integrals over $\eta_3$ and $\eta_4$ in the same manner as in the \eqref{integral}, one obtains that
\begin{equation} \label{type2}
    G^K_{b} (\eta_1, x_1| \eta_2,x_2)  \sim \lambda^2 \eta, \quad {\rm as} \quad  \eta \to +\infty.
\end{equation}
For the third type of diagrams that are depicted on the Fig. \ref{tadpole4} (again, denoting their contribution as $G^K_{c}$) in the limit \eqref{limit} one finds:
\begin{multline} \label{type3}
    G^K_{c} (\eta_1, x_1| \eta_2,x_2) = - \frac{\lambda^2}{4} \int^{\infty}_{\eta_0} d \eta_3\, dx_3\, C(\eta_3) \int^{\infty}_{\eta_0} d \eta_4\, dx_4\, C(\eta_4) \, \bigg(  G^K_{13} G^K_{33} G^A_{34} G^{K}_{44} G^A_{42} + \\+G^R_{13} G^K_{33} G^K_{34} G^K_{44} G^A_{42} + G^R_{13} G^K_{33} G^R_{34} G^K_{44} G^K_{42} \bigg) \sim \lambda^2 \eta^2, \quad {\rm as} \quad \eta \to +\infty.
\end{multline}
Thus, from \eqref{type1}, \eqref{type2}, and \eqref{type3}, one can see that even if $\lambda$ is very small the loop corrections to the Keldysh propagator can become large after a long enough time of evolution. It means that to understand the physics one has to resum at least the leading contributions from all loops. We will see in a moment that growing as $\eta = (\eta_1 + \eta_2)/2 \to \infty$ contributions from tadpoles can be absorbed into the mass and mode renormalization, because products of the modes $g_p(\eta_1)\, g^*_p(\eta_2)$ contain $\eta$ dependent terms in the future infinity region. 

At this moment note that the second type of diagrams from the Fig.\ref{tadpole4} is suppressed by the additional power of $\lambda$. Hence, to resum the leading contributions one just needs to take into account the chains of bubbles and ignore the ``cactus'' type of diagrams. These observations enable us to write the Dyson-Schwinger equation for the resummation of the tadpole corrections to the Keldysh propagator. Denoting the ``exact'' (resummed) Keldysh propagator as $\Tilde{G}^K$, we find that it obeys the equation as follows:
\begin{equation}
    \Tilde{G}^K_{12} = G^K_{12} -\frac{i\lambda}{2} \int^{\infty}_{\eta_0} d \eta_3\, dx_3\, C(\eta_3) \, \bigg( G^K_{13} {G}^K_{33} \Tilde{G}^A_{32} + G^R_{13} {G}^K_{33} \Tilde{G}^K_{32} \bigg).
\end{equation}
Here $\Tilde{G}^A_{32}$ is the advanced Green function with the resummed bubble diagrams, which is obtained with the use of (\ref{G_R_A_tadpole}).

Applying the differential operator $(\Box+m^2)$ to the both sides of this equation, one obtains
\begin{equation} \label{tad_eq}
    \bigg( \Box +m^2+\frac{\lambda}{2} {G}^K_{11} \bigg) \Tilde{G}^K_{12}=0.
\end{equation}
The equations for the exact retarded and advanced propagators have the same form with the delta-functions on the right hand side. 
%Hence, one can see that if the temporal part of the modes was just single exponential, $g_k(\eta) \sim e^{-i\w_k \eta}$, for all times, then the tadpole diagrams just lead to the mass renormalization as the $\Tilde{G}^K_{11}$ term is just the divergence of the propagator in the coincident points:
Thus, one can take into account the leading tadpole diagrams as \textcolor{black}{a sort of a ``mass renormalization''}
\begin{equation} \label{mass_ren}
    \Tilde{m}^2 = m^2 + \frac{\lambda}{2} {G}^K_{11}.
\end{equation}
\textcolor{black}{There are certain comments}, which are in order at this point because the ``mass renormalization'' seem to depend here on space-time coordinates. 

In fact, let us consider the tree-level Keldysh propagator $G^K$ at the coinciding points. For simplicity we consider in detail the propagator for the Fock space ground state\footnote{The situation with other states with {\it zero} anomalous averages is not much different.}:
\begin{equation} \label{g11}
    {G}^K_{11} = \int dk \, g^{in}_k(\eta_1) g^{in*}_k(\eta_1).
\end{equation}
Obviously, ${G}^K_{11}$ does not depend on the coordinates in the remote past, $\eta_1 \to -\infty$, as the modes $g^{in}_k(\eta_1)$ in this region of space-time are given by simple exponents \eqref{in_modes_asympt}, and the integral over $k$ just gives the standard UV divergence due to zero-point fluctuations: 
\begin{equation} \label{gg1}
    \int dk \, g^{in}_k(\eta_1) g^{in*}_k(\eta_1) \xrightarrow[\eta_1 \to -\infty]{} \int \frac{dk}{4\pi \w_{in}}.
\end{equation}
Now let us concentrate on the opposite limit $\eta_1 \to +\infty$. Plugging asymptotic behavior of the in-modes \eqref{in_modes_asympt} into \eqref{g11}, one obtains: \textcolor{black}{
\begin{multline} \label{gg2}
    \int dk \, g^{in}_k(\eta_1) g^{in*}_k(\eta_1) \xrightarrow[\eta_1 \to +\infty]{} \int \frac{dk}{4\pi \w_{in}} \bigg[ |\alpha|^2 + |\beta|^2 + \alpha \beta^* \, e^{-2i \w_{out} \eta_1} + \alpha^* \beta \, e^{2i\w_{out} \eta_1}  \bigg] \approx \\ \approx \int \frac{dk}{4\pi \w_{in}} \bigg[ |\alpha|^2 + |\beta|^2 \bigg] = \int \frac{dk}{4\pi \w_{out}} \frac{\sinh^2 (\pi \w_+/\rho) + \sinh^2 (\pi \w_- / \rho)}{\sinh(\pi \w_{in} / \rho) \sinh (\pi \w_{out} / \rho)},
\end{multline}}
where we have dropped the rapidly oscillating terms as their contribution to the integral is negligible in comparison with the contribution of the terms that are independent of time, $\eta_1$.

Now, using \eqref{tad_eq}, we can write down the equation for the tadpole corrected modes $\Tilde{g}_k(\eta)$ as:
\begin{equation}
   \bigg[ \frac{d^2}{d\eta^2} + k^2 + C(\eta) \, \bigg( m^2 + \frac{\lambda}{2} \int dq \, g_q^{in}(\eta) g_q^{in*}(\eta)  \bigg) \bigg] \Tilde{g}_k(\eta) =0.
\end{equation}
\textcolor{black}{The term in this expression which is proportional to $\lambda$ contains the standard UV divergence which can be absorbed in to the renormalization of $m$. But also this term contains finite $\eta$ dependent contributions. To treat them we consider the changes in the modes.} In fact, using the asymptotic forms \eqref{gg1} and \eqref{gg2}, one can find that the in-modes have the following renormalized asymptotic behavior:
\begin{align} \label{tad_in_asympt}
    \Tilde{g}_k^{in}(\eta) \approx
    \begin{cases}
        \frac{1}{\sqrt{4\pi \Tilde{\w}_{in}}} e^{-i\Tilde{\w}_{in}\eta}, \qquad  &\text{as} \quad \eta \to -\infty \\
        \frac{1}{\sqrt{4\pi \Tilde{\w}_{in}}} \bigg[ {C}_1(k) \,  e^{-i\Tilde{\w}_{out}\eta} + {C}_2(k) \, e^{i\Tilde{\w}_{out}\eta} \bigg], \qquad &\text{as} \quad \eta \to +\infty,
    \end{cases}
\end{align}
where to find the coefficients ${C}_{1,2}(k)$ one needs to calculate the integral $\int dq \, g_q^{in}(\eta) g_q^{in*}(\eta)$ explicitly to know the potential in the equation under consideration. To the best of our knowledge this integral is not a table one, but below we will not use the explicit form of ${C}_{1,2}(k)$. 

At the same time, $\Tilde{\w}_{in}$ and $\Tilde{\w}_{out}$ are given by
\begin{align} 
    &\Tilde{\w}_{in}^2(k) \approx k^2 + (A-B) \bigg( m^2 + \int \frac{dk}{4\pi \w_{in}} \bigg), \nonumber \\
    &\Tilde{\w}_{out}^2(k) \approx k^2 + (A+B) \bigg( m^2 + \int \frac{dk}{4\pi \w_{out}} \frac{\sinh^2 (\pi \w_+/\rho) + \sinh^2 (\pi \w_- / \rho)}{\sinh(\pi \w_{in} / \rho) \sinh (\pi \w_{out} / \rho)} \bigg). \label{tilde_omega}
\end{align}
Then, using the tadpole corrected in-modes $\Tilde{g}^{in}_k(\eta)$ one can construct the Keldysh propagator $\Tilde{G}^K$ according to the eq. \eqref{gk_modes}. For example, in the Fock space ground state for the in-modes one has
\begin{equation}
    \Tilde{G}^K(\eta,x|\eta',x') = \int^{\infty}_{-\infty} dk \, \frac{1}{2} \Big[ \Tilde{u}_k^{in}(\eta,x) \, \Tilde{u}_k^{in*}(\eta',x') + \Tilde{u}_k^{in}(\eta',x') \, \Tilde{u}_k^{in*}(\eta,x) \Big], \quad {\rm where} \quad \Tilde{u}_k^{in}(\eta,x) = e^{ikx} \Tilde{g}_k^{in}(\eta).
\end{equation}
In the same manner, we can obtain the asymptotic behavior of the out-modes
\begin{align} \label{tad_out_asympt}
    \Tilde{g}_k^{out}(\eta) \approx
    \begin{cases}
        \frac{1}{\sqrt{4\pi \Tilde{\w}_{in}}} \bigg[ {C}_3(k) \,  e^{-i\Tilde{\w}_{in}\eta} + {C}_4(k) \, e^{i\Tilde{\w}_{in}\eta} \bigg] , \qquad  &\text{as} \quad \eta \to -\infty \\
        \frac{1}{\sqrt{4\pi \Tilde{\w}_{in}}} e^{-i\Tilde{\w}_{out}\eta} , \qquad &\text{as} \quad \eta \to +\infty,
    \end{cases}
\end{align}
and construct the tadpole corrected propagator using these modes.

In all, we obtain that the sum of the leading tadpole diagrams in the limit \eqref{limit} results in the change of modes from $g_k(\eta)$ (with the asymptotic behavior of which is given in \eqref{in_modes_asympt} and \eqref{out_modes_asympt}) to $\Tilde{g}_k(\eta)$ (with the asymptotic behavior of which is given in \eqref{tad_in_asympt} and \eqref{tad_out_asympt}). And all three propagators change accordingly in the same way. In the following we assume that the modes (and propagator which are constructed from these modes) are tadpole corrected, i.e. are given by \eqref{tad_in_asympt} and \eqref{tad_out_asympt}. However, to reduce notations, we will omit the tilde symbol for these modes and propagators. \textcolor{black}{Also, we assume that the UV divergence of the form $\int \frac{dk}{4\pi \w_{in}}$ is absorbed into the renormalization of the bare mass $m$, and the tadpole corrected frequencies $\Tilde{\w}_{in}$ and $\Tilde{\w}_{out}$ in \eqref{tilde_omega} are finite.}

\subsection{Sunset diagrams}
%If we restrict ourselves to the one- and two-loop level, there are two types of diagrams that arise in the loop contributions to the propagators: tadpole and sunset diagrams. The sum of the tadpole diagrams is treated in the Appendix \ref{tadpole}, where we show that for a state with the zero initial anomalous quantum average these loop corrections correspond to a mass renormalization. %\textcolor{red}{Надо бы добавить картинку tadpole диаграмм.}

Let us continue with the consideration of the two-loop sunset diagrams which lead to more physically interesting contributions to the two-point functions. Namely one cannot absorb their contribution into the mass and wave function renormalization. 

Let us start with the consideration of the sunset contribution to the retarded propagator. The sunset diagrams lead to the following corrections of the retarded propagator:
\begin{multline}
    G^R_{(2)}(\eta_1, x_1| \eta_2,x_2) = - \frac{\lambda^2}{24} \int^{\infty}_{\eta_0} d \eta_3\, dx_3\, C(\eta_3) \int^{\infty}_{\eta_0} d \eta_4\, dx_4\, C(\eta_4) \, \bigg( 6 G^R_{13} G^R_{34} (G^K_{34})^2 G^R_{42} + \\+G^R_{13} (G^R_{34})^2 G^R_{42} \bigg) =- \frac{\lambda^2}{24} \int^{\infty}_{\eta_0} d \eta_3\, dx_3\, C(\eta_3)\,  \int^{\infty}_{\eta_0} d \eta_4\, dx_4\, C(\eta_4) \,\theta(\eta_1-\eta_3) \theta(\eta_3-\eta_4) \theta(\eta_4-\eta_2) \times \\ \times \bigg( 6 G_{13} G_{34} (G^K_{34})^2 G_{42} + G_{13} (G_{34})^2 G_{42} \bigg).
\end{multline}
%\textcolor{red}{Странно, что для кубической теории ты не обсуждал запаздывающий пропагатор, а для квартичной теперь обсуждаешь. Это вызовет недоразумения. Я добавлю комментарии в обсуждении кубической теории.}
In the limit \eqref{limit}, this two-loop contribution contains the multiplier, which is approximately equal to $\theta(\eta-\eta_3) \theta(\eta_3-\eta_4) \theta(\eta_4-\eta)=0$. The multiplier appears due to the causal nature of the retarded Green function \cite{Kamenev}. Because of this fact there are no growing with average time $\eta$ two-loop contributions to the retarded propagator. That is the usual story in the limit that we consider \cite{Akhmedov:2013vka}. Similarly one can show that for the same reason advanced propagator does not receive secularly growing contributions in the same limit. Meanwhile, as we will see in a moment, Keldysh propagator does receive growing with time contributions from the sunset diagrams. This is the key difference of the sunset contributions as compared to the tadpole ones. Because of that sunset corrections cannot be attributed to the mass renormalization. Their growth in time is due to the change of the state of the theory.

\begin{figure}[h]
\includegraphics[width=15cm]{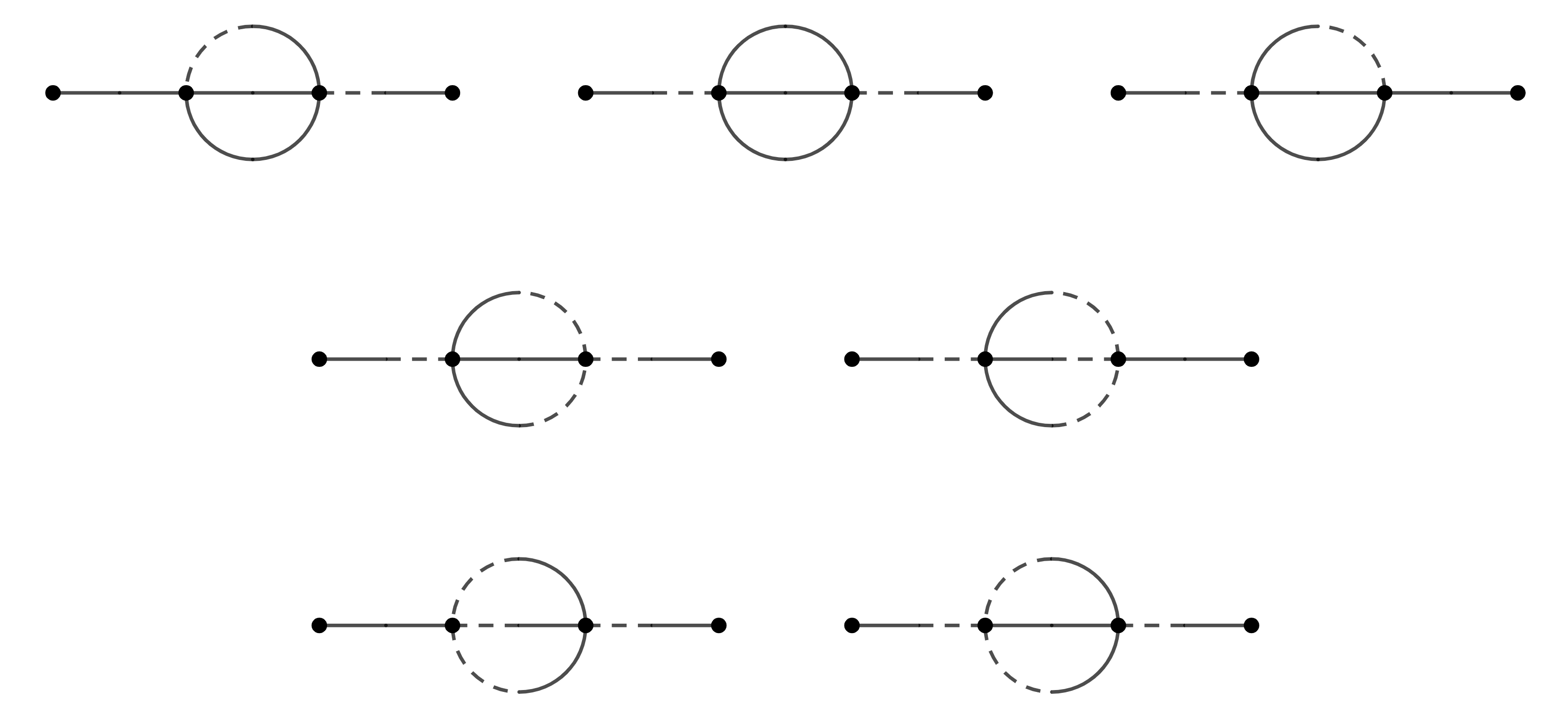}
\centering
\caption{Sunset diagrams providing the two loop correction to the Keldysh propagator.}
\label{sunset_4}
\end{figure}

For the Keldysh propagator the sunset diagrams are depicted on the Fig. \ref{sunset_4}. Then, the two loop correction to the Keldysh propagator has the following form:
\begin{multline} \label{K2}
G^K_{(2)}(\eta_1,x_1|\eta_2,x_2) = -\frac{\lambda^2}{6} \int_{\eta_0}^{\infty} d\eta_3 \, dx_3 C(\eta_3) \int_{\eta_0}^{\infty} d\eta_4\, dx_4 C(\eta_4) \bigg[ 3 G^K_{13} (G^K_{34})^2 G^A_{34} G^A_{42} + G^R_{13} (G^K_{34})^3 G^A_{42} + \\+3 G^R_{13} G^R_{34} (G^K_{34})^2 G^K_{42} 
+ \frac{3}{4} G^R_{13} (G^R_{34})^2 G^K_{34} G^A_{42} + \frac{1}{4} G^R_{13} (G^R_{34})^3  G^K_{42} +\frac{1}{4} G^K_{13} (G^A_{34})^3  G^A_{42} + \frac{3}{4} G^R_{13} (G^A_{34})^2 G^K_{34}  G^A_{42} \bigg].
\end{multline}
%and comparing this expression with \eqref{keldysh_generic}, one can distinguish here the loop corrections to the occupation number and anomalous quantum average. For the occupation number in the limit \eqref{limit} one has
We start with the consideration of the in-modes \eqref{tad_in_asympt} and want to investigate how do the two-loop corrections to the Keldysh propagator \eqref{K2} affect the initial occupation number and anomalous quantum average. Namely, we consider a spatially homogeneous initial state with the planckian occupation number and with the zero anomalous quantum average:
\begin{equation}\label{initailstate}
    \langle a^{\dagger}_q a_k \rangle = n_k\, \delta(k-q), \quad n_k = \frac{1}{e^{\omega_{in}/T} - 1}, \quad {\rm and} \quad \langle a_q a_k \rangle = 0.
\end{equation}
Here $T$ is some temperature.

From the expression \eqref{K2} we can extract loop corrections to the occupation number and anomalous quantum average by comparing it with eq. \eqref{keldysh_generic} in the limit \eqref{limit}. Then for the occupation number one obtains the expression as follows:
\begin{multline}
    n_p^{(2)} \approx -(2\pi)^2\frac{\lambda^2}{6} \int^{\eta}_{\eta_0}d\eta_3 C(\eta_3) \int^{\eta}_{\eta_0}d\eta_4 C(\eta_4) \int dq\,dr\,ds\, \delta(p-q-r-s) g_p^{in*}(\eta_3) g_p^{in}(\eta_4) \times \\ \times
    \bigg[ \bigg( \frac{1}{2} + n_p \bigg) \bigg( 3G^K_{q34} G^K_{r34} G_{s34} + \frac{1}{4} G_{q34}G_{r34}G_{s34} \bigg) - G^K_{q34} G^K_{r34} G^K_{s34} - \frac{3}{4} G_{q34} G_{r34} G^K_{s34} \bigg],
\end{multline}
where we have neglected in the leading approximation the difference between $\eta_1, \eta_2$ and $\eta = (\eta_1 + \eta_2)/2$ in the limit under consideration.

Expanding propagators in mode functions, one obtains that
\begin{multline} \label{n_phi4_modes}
    n_p^{(2)} \approx (2\pi)^2 \frac{\lambda^2}{6} \int^{\eta}_{\eta_0}d\eta_3 C(\eta_3) \int^{\eta}_{\eta_0}d\eta_4 C(\eta_4) \int dq\,dr\,ds\, \delta(p-q-r-s) g_p^{in*}(\eta_3) g_p^{in}(\eta_4) \times \\ \times
    \bigg[   g_q^{in*}(\eta_3)g_q^{in}(\eta_4) g_r^{in*}(\eta_3)g_r^{in}(\eta_4) g_s^{in*}(\eta_3)g_s^{in}(\eta_4) \bigg( (1+n_p)(1+n_{q})(1+n_{r})(1+n_{s}) - n_p n_{q} n_{r} n_{s} \bigg) +\\
    + 3g_q^{in*}(\eta_3)g_q^{in}(\eta_4) g_r^{in*}(\eta_3)g_r^{in}(\eta_4) g_s^{in}(\eta_3)g_s^{in*}(\eta_4) \bigg( (1+n_p)(1+n_{q})(1+n_{r})n_s - n_p n_{q} n_{r} (1+n_s) \bigg)+\\
    +3g_q^{in*}(\eta_3)g_q^{in}(\eta_4) g_r^{in}(\eta_3)g_r^{in*}(\eta_4) g_s^{in}(\eta_3)g_s^{in*}(\eta_4) \bigg( (1+n_p)(1+n_{q})n_r n_s - n_p n_{q} (1+n_r) (1+n_s) \bigg)+\\
    + g_q^{in}(\eta_3)g_q^{in*}(\eta_4) g_r^{in}(\eta_3)g_r^{in*}(\eta_4) g_s^{in}(\eta_3)g_s^{in*}(\eta_4) \bigg( (1+n_p)n_q n_r n_s - n_p (1+n_q)(1+n_r)(1+n_s) \bigg)
    \bigg].
\end{multline}
To estimate the time integrals in this expression we perform the same operations as in the previous section, i.e. we divide the integration interval over temporal coordinates $\eta_{3,4}$ into the same three regions as in the previous section and use the asymptotic behavior of the modes \eqref{tad_in_asympt} in the remote past and future. Intermediate region cannot lead to growing contributions due to its fixed length even if $\eta_0$ and $\eta$ are taken to the past and future infinities, correspondingly. 

In (\ref{n_phi4_modes}) there are three virtual momenta, and if the delta function has two frequencies with one sign and other two with the opposite sign (e.g. $\delta(\w_{out}(p) + \w_{out}(q) - \w_{out}(r) - \w_{out}(s))$), then the argument of the delta-function can be zero. In such a case we have a growing with time $\eta$ contribution to (\ref{n_phi4_modes}). Evaluating integrals over $\eta_3$ and $\eta_4$ and keeping only those terms, which in the limit \eqref{limit} give the delta-functions of the form $\delta(\w_{out}(p) + \w_{out}(q) - \w_{out}(r) - \w_{out}(s))$ (i.e. the terms that are independent of $t=(\eta_3+\eta_4)/2$ under the integral), one obtains
\begin{multline} \label{n_p_prefin}
    n_p^{(2)} \approx \frac{\lambda^2(A+B)^2 \, \eta}{64 \pi} \int \frac{dq \, dr \, ds \, \delta\Big(p-q-r-s\Big)}{\w_{in}(p)\w_{in}(q)\w_{in}(r) \w_{in}(s)} \delta\Big(\w_{out}(p) + \w_{out}(q) - \w_{out}(r) - \w_{out}(s)\Big) \times
     \\ \times \bigg\{  \bigg( | C_1 (p) C_1(q) C_2(r) C_2(s)|^2+|C_2(p) C_2(q) C_1(r) C_1(s) |^2 \bigg)
    \Big[ (1+n_p)(1+n_{q})(1+n_{r})(1+n_{s}) - n_p n_{q} n_{r} n_{s} \Big] + \\+2 \bigg( |C_1(p) C_1(q) C_2(r) C_1(s)|^2+|C_2(p) C_2(q) C_1(r) C_2(s) |^2 \bigg)\Big[ (1+n_p)(1+n_{q})(1+n_{r})n_s - n_p n_{q} n_{r} (1+n_s) \Big] + \\+ \bigg( | C_1(p)  C_2(q) C_2(r) C_2(s)|^2  +|C_2(p) C_1(q) C_1(r) C_1(s)|^2 \bigg) \Big[ (1+n_p)n_q(1+n_{r})(1+n_{s}) - n_p (1+n_q) n_{r}  n_{s}  \Big] + \\+ \bigg( |C_1(p) C_1(q) C_1(r) C_1(s)|^2+|C_2(p) C_2(q) C_2(r) C_2(s) |^2 \bigg) \Big[ (1+n_p)(1+n_{q})n_r n_s - n_p n_{q} (1+n_r)(1+n_s) \Big] + \\+2 \bigg( |C_1(p) C_2(q) C_1(r) C_2(s)|^2+|C_2(p) C_1(q) C_2(r) C_1(s)|^2 \bigg) \Big[ (1+n_p)n_q n_r (1+n_{s})  - n_p (1+n_q) (1+n_r) n_{s}\Big] + \\+ \bigg(|C_1(p) C_2(q) C_1(r) C_1(s)|^2+|C_2(p) C_1(q) C_2(r) C_2(s) |^2 \bigg)\Big[ (1+n_p)n_q n_r n_s - n_p (1+n_q)(1+n_r)(1+n_s) \Big] \bigg\}.
\end{multline}
In writing this expression we have neglected $\Bar{\eta}'$ with respect to $\eta \to \infty$.

However, one should note that the presence of the delta-functions $\delta(p-q-r-s)$ and 
$\delta(\w_{out}(p) + \w_{out}(q) - \w_{out}(r) - \w_{out}(s))$ restricts the momenta to satisfy the following system of equations
\begin{align} \label{mom_sol1}
    \begin{cases}
        p=q+r+s,\\
        \w_{out}(p) + \w_{out}(q) = \w_{out}(r) + \w_{out}(s).
    \end{cases}
\end{align}
This system has the following solutions
\begin{align} \label{mom_sol2}
    \begin{cases}
        r=p\\
        s=-q
    \end{cases}
    \quad \text{and} \qquad
    \begin{cases}
        r=-q\\
        s=p
    \end{cases},
\end{align}
which is the specifics of the 2D kinematics.

Plugging these solutions into \eqref{n_p_prefin}, one can see that the term proportional to $\Big[(1+n_p)(1+n_{q})n_r n_s - n_p n_{q} (1+n_r)(1+n_s)\Big]$ vanishes identically. In 2D space-time this happens for any distribution, while in higher dimensions only for the thermal (planckian) one. Finally, the expression for the two-loop correction to the occupation number has the following form
%\begin{multline} \label{n_p_fin}
%    n_p^{(2)} \approx \frac{\lambda^2(A+B)^2(\eta-\Bar{\eta}')}{64 \pi} \int \frac{dq \, dr \, ds \, \delta\Big(p-q-r-s\Big)}{\w_{in}(p)\w_{in}(q)\w_{in}(r) \w_{in}(s)} \delta\Big(\w_{out}(p) + \w_{out}(q) - \w_{out}(r) - \w_{out}(s)\Big) \times
%     \\ \times \bigg\{  \bigg( | C_1 (p) C_1(q) C_2(r) C_2(s)|^2+|C_2(p) C_2(q) C_1(r) C_1(s) |^2 \bigg)
%    \Big[ (1+n_p)(1+n_{q})(1+n_{r})(1+n_{s}) - n_p n_{q} n_{r} n_{s} \Big] + \\+2 \bigg( |C_1^*(p) C_1^*(q) C_2^*(r) C_1(s)|^2+|C_2^*(p) C_2^*(q) C_1^*(r) C_2(s) |^2 \bigg)\Big[ (1+n_p)(1+n_{q})(1+n_{r})n_s - n_p n_{q} n_{r} (1+n_s) \Big] + \\+ \bigg( | C_1^*(p)  C_2(q) C_2^*(r) C_2^*(s)|^2  +|C_2^*(p) C_1(q) C_1^*(r) C_1^*(s)|^2 \bigg) \Big[ (1+n_p)n_q(1+n_{r})(1+n_{s}) - n_p (1+n_q) n_{r}  n_{s}  \Big] + \\+2 \bigg( |C_1^*(p) C_2(q) C_1(r) C_2^*(s)|^2+|C_2^*(p) C_1(q) C_2(r) C_1^*(s)|^2 \bigg) \Big[ (1+n_p)n_q n_r (1+n_{s})  - n_p (1+n_q) (1+n_r) n_{s}\Big] + \\+ \bigg(|C_1^*(p) C_2(q) C_1(r) C_1(s)|^2+|C_2^*(p) C_1(q) C_2(r) C_2(s) |^2 \bigg)\Big[ (1+n_p)n_q n_r n_s - n_p (1+n_q)(1+n_r)(1+n_s) \Big] \bigg\}.
%\end{multline}

\begin{mybox}
\begin{align}  
    n_p^{(2)} \approx \frac{\lambda^2(A+B)^2 \, \eta}{64 \pi} \int \frac{dq \, dr \, ds \, \delta\Big(p-q-r-s\Big)}{\w_{in}(p)\w_{in}(q)\w_{in}(r) \w_{in}(s)} \delta\Big(\w_{out}(p) + \w_{out}(q) - \w_{out}(r) - \w_{out}(s)\Big) \times \nonumber
     \\ \times \bigg\{  \mathcal{N}_1 (p,q,r,s)
    \Big[ (1+n_p)(1+n_{q})(1+n_{r})(1+n_{s}) - n_p n_{q} n_{r} n_{s} \Big] + \nonumber \\+\mathcal{N}_2(p,q,r,s) \Big[ (1+n_p)(1+n_{q})(1+n_{r})n_s - n_p n_{q} n_{r} (1+n_s) \Big] + \nonumber\\+ \mathcal{N}_3(p,q,r,s) \Big[ (1+n_p)n_q(1+n_{r})(1+n_{s}) - n_p (1+n_q) n_{r}  n_{s}  \Big] +\nonumber \\+\mathcal{N}_4(p,q,r,s) \Big[ (1+n_p)n_q n_r (1+n_{s})  - n_p (1+n_q) (1+n_r) n_{s}\Big] +\nonumber\\+ \mathcal{N}_5(p,q,r,s) \Big[ (1+n_p)n_q n_r n_s - n_p (1+n_q)(1+n_r)(1+n_s) \Big] \bigg\}, \label{n_p_fin}
\end{align}
\end{mybox}
where the quantities $\mathcal{N}_{1,2,3,4,5}(p,q,r,s)$ are given by:
\begin{align}
    &\mathcal{N}_1(p,q,r,s) = | C_1 (p) C_1(q) C_2(r) C_2(s)|^2+|C_2(p) C_2(q) C_1(r) C_1(s) |^2, \nonumber \\
    &\mathcal{N}_2(p,q,r,s) = 2 \bigg( |C_1(p) C_1(q) C_2(r) C_1(s)|^2+|C_2(p) C_2(q) C_1(r) C_2(s) |^2 \bigg),\nonumber \\
    &\mathcal{N}_3(p,q,r,s) = | C_1(p)  C_2(q) C_2(r) C_2(s)|^2  +|C_2(p) C_1(q) C_1(r) C_1(s)|^2,\nonumber \\
    &\mathcal{N}_4(p,q,r,s) = 2 \bigg( |C_1(p) C_2(q) C_1(r) C_2(s)|^2+|C_2(p) C_1(q) C_2(r) C_1(s)|^2 \bigg),\nonumber \\
    &\mathcal{N}_5(p,q,r,s) = |C_1(p) C_2(q) C_1(r) C_1(s)|^2+|C_2(p) C_1(q) C_2(r) C_2(s) |^2.
\end{align}
Let us continue now with the corrections to the anomalous quantum average:
\begin{multline} \label{kappa_2}
    \kappa_p^{(2)} \approx (2\pi)^2\frac{\lambda^2}{6} \int^{\eta}_{\eta_0}d\eta_3 \, C(\eta_3) \int^{\eta}_{\eta_0}d\eta_4 \, C(\eta_4) \int dq\,dr\,ds\, \delta\Big(p-q-r-s\Big) \,  g_p^{in*}(\eta_3) \, g_p^{in*}(\eta_4) \times \\ \times
    \bigg\{ 2 \theta(\eta_4-\eta_3)\bigg( \frac{1}{2} + n_p \bigg) \bigg( 3G^K_{q34} G^K_{r34} G_{s34} + \frac{1}{4} G_{q34}G_{r34}G_{s34} \bigg) - G^K_{q34} G^K_{r34} G^K_{s34} - \frac{3}{4} G_{q34} G_{r34} G^K_{s34} \bigg\}.
\end{multline}
Expanding in this expression the propagators in the modes and making the exchange $\eta_3 \leftrightarrow \eta_4$ in some of the integrals, one can rewrite the last expression as
\begin{multline} \label{kappa_4_modes}
    \kappa_p^{(2)} \approx -(2\pi)^2\frac{\lambda^2}{6} \int^{\eta}_{\eta_0}d\eta_3 C(\eta_3) \int^{\eta}_{\eta_0}d\eta_4 C(\eta_4) \int dq\,dr\,ds\, \delta\Big(p-q-r-s\Big) \, g_p^{in*}(\eta_3) \, g_p^{in*}(\eta_4) \times \\ \times
    \bigg\{ g^{in*}_q(\eta_3) g_q^{in}(\eta_4) g^{in*}_r(\eta_3) g_r^{in}(\eta_4) g^{in*}_s(\eta_3) g_s^{in}(\eta_4) \bigg[ (1+2n_p) \bigg( (1+n_q)(1+n_r)(1+n_s)-n_q n_r n_s \bigg) \times \\ \times \bigg( \theta(\eta_4-\eta_3 ) - \theta(\eta_3-\eta_4) \bigg) + \bigg( (1+n_q)(1+n_r)(1+n_s)+n_q n_r n_s \bigg)  \bigg] +\\+
    3 g^{in*}_q(\eta_3) g_q^{in}(\eta_4) g^{in*}_r(\eta_3) g_r^{in}(\eta_4) g_s^{in}(\eta_3) g^{in*}_s(\eta_4) \bigg[ (1+2n_p) \bigg( (1+n_q)(1+n_r)n_s - n_q n_r (1+n_s)  \bigg) \times \\ \times \bigg( \theta(\eta_4-\eta_3 ) - \theta(\eta_3-\eta_4) \bigg) +\bigg((1+n_q)(1+n_r)n_s + n_q n_r (1+n_s)  \bigg) \bigg]
    \bigg\}.
\end{multline}
Here the terms that do not contain theta-functions give delta-functions after integrating over $\eta_3$ and $\eta_4$. In this case the situation is similar to the one with the occupation number \eqref{n_p_fin}. The terms that do contain theta-functions give contributions in which the internal momenta are not restricted by the energy conservation:
\begin{equation} \label{int_theta}
    \int^{\eta-\Bar{\eta}'}_{\Bar{\eta}'-\eta} d\tau \, e^{i\tau \w} \bigg( \theta(-\tau ) - \theta(\tau) \bigg) = - \frac{2i}{\w} \bigg( 1-\cos \big(\w(\eta - \Bar{\eta}') \big) \bigg) \to - \frac{2i}{\w}, \qquad \eta \to \infty.
\end{equation}
%Here we have used the fact that
%\begin{equation}
%    \int^{\infty}_{-\infty} \frac{\cos (\w \eta)}{\w} f(\w) d\w \approx  f(0)  \int^{\infty}_{-\infty} \frac{\cos (\w) }{\w}  d\w \myeq{} 0, \qquad \eta \to \infty.
%\end{equation}
Then, the two-loop contribution to the anomalous quantum average can be written as
%\begin{equation} \label{kappa_4_fin}
%    \kappa^{(2)}_p \approx (\eta-\Bar{\eta}') \big[  I_1(p,q,r,s) + I_2(p,q,r,s) \big],
%\end{equation}
\begin{mybox}
    \begin{align}  
    \kappa^{(2)}_p \approx - \frac{\lambda^2 (A+B)^2 \, \eta}{64\pi} &\int \frac{dq \, dr \, ds \, \delta\Big(p-q-r-s\Big)}{\w_{in}(p)\w_{in}(q)\w_{in}(r) \w_{in}(s)} \times \nonumber \\ \times \bigg\{ &\mathcal{K}_1(p,q,r,s) \Big[ (1+n_q)(1+n_r)(1+n_s)+n_q n_r n_s  \Big] + \nonumber \\+ &\mathcal{K}_2(p,q,r,s) \Big[ (1+n_q)(1+n_r)n_s+n_q n_r (1+n_s)  \Big] +\nonumber \\ +&\mathcal{K}_3(p,q,r,s) \Big[ n_q(1+n_r)(1+n_s)+(1+n_q) n_r n_s  \Big] +\nonumber \\+ &\mathcal{K}_4(p,q,r,s) (1+2n_p)\Big[ (1+n_q)(1+n_r)(1+n_s)-n_q n_r n_s  \Big] +\nonumber \\+ &\mathcal{K}_5(p,q,r,s) (1+2n_p) \Big[ (1+n_q)(1+n_r)n_s-n_q n_r (1+n_s)  \Big] \bigg\}  ,\label{kappa_4_fin}
\end{align}
\end{mybox}
%where $I_1(p,q,r,s)$ and $I_2(p,q,r,s)$ are given by
where the quantities $\mathcal{K}_{1,2,3,4,5}(p,q,r,s)$ are given by:

\begin{align*}
    \mathcal{K}_1(p,q,r,s) &= \bigg( C_1^*(p) C_1^*(q) C_2^*(r) C_2^*(s) C_2^*(p) C_1(q) C_2(r) C_2(s) + \nonumber \\ +&C_2^*(p) C_2^*(q) C_1^*(r) C_1^*(s) C_1^*(p) C_2(q) C_1(r) C_1(s) \bigg) \delta\Big(\w_{out}(p) + \w_{out}(q) - \w_{out}(r) - \w_{out}(s)\Big),\nonumber \\
    \mathcal{K}_2(p,q,r,s) &= 2 \bigg( C_1^*(p) C_1^*(q) C_2^*(r) C_1(s) C_2^*(p) C_1(q) C_2(r) C_1^*(s) +\nonumber \\ +&C_2^*(p) C_2^*(q) C_1^*(r) C_2(s) C_1^*(p) C_2(q) C_1(r) C_2^*(s) \bigg)\delta\Big(\w_{out}(p) + \w_{out}(q) - \w_{out}(r) - \w_{out}(s)\Big),\nonumber \\
    \mathcal{K}_3(p,q,r,s) &= \bigg( C_1^*(p) C_2(q) C_2^*(r) C_2^*(s) C_2^*(p) C_2^*(q) C_2(r) C_2(s) +\nonumber \\+  &C_2^*(p) C_1(q) C_1^*(r) C_1^*(s) C_1^*(p) C_1^*(q) C_1(r) C_1(s) \bigg) \bigg)\delta\Big(\w_{out}(p) + \w_{out}(q) - \w_{out}(r) - \w_{out}(s)\Big),
\end{align*}
and
\begin{multline*}
    \mathcal{K}_4(p,q,r,s) = \\=- \frac{i}{3\pi} \bigg( \frac{C_1^*(p) C_1^*(q) C_1^*(r) C_1^*(s) C_2^*(p) C_1(q) C_1(r) C_1(s) - C_2^*(p) C_2^*(q) C_2^*(r) C_2^*(s) C_1^*(p) C_2(q) C_2(r) C_2(s) }{\w_{out}(p) + \w_{out}(q) + \w_{out}(r) + \w_{out}(s)} + \\+ 3 \frac{C_1^*(p) C_1^*(q) C_1^*(r) C_2^*(s) C_2^*(p) C_1(q) C_1(r) C_2(s) - C_2^*(p) C_2^*(q) C_2^*(r) C_1^*(s) C_1^*(p) C_2(q) C_2(r) C_1(s)  }{\w_{out}(p) + \w_{out}(q) + \w_{out}(r) - \w_{out}(s)} + \\+ 3 \frac{ C_1^*(p) C_1^*(q) C_2^*(r) C_2^*(r) C_2^*(p) C_1(q) C_2(r) C_2(s) - C_2^*(p) C_2^*(q) C_1^*(r) C_1^*(s) C_1^*(p) C_2(q) C_1(r) C_1(s) }{\w_{out}(p) + \w_{out}(q) - \w_{out}(r) - \w_{out}(s)} + \\+ \frac{ C_1^*(p) C_2^*(q) C_2^*(r) C_2^*(s) C_2^*(p) C_2(q) C_2(r) C_2(s) - C_2^*(p) C_1^*(q) C_1^*(r) C_1^*(s) C_1^*(p) C_1(q) C_1(r) C_1(s) }{\w_{out}(p) - \w_{out}(q) - \w_{out}(r) - \w_{out}(s)} \bigg),
\end{multline*}
\begin{multline*}
    \mathcal{K}_5(p,q,r,s) = \\=- \frac{i}{\pi} \bigg( \frac{ C_1^*(p) C_1^*(q) C_1^*(r) C_2(s) C_2^*(p) C_1(q) C_1(r) C_2^*(s) - C_2^*(p) C_2^*(q) C_2^*(r) C_1(s) C_1^*(p) C_2(q) C_2(r) C_1^*(s) }{\w_{out}(p) + \w_{out}(q) + \w_{out}(r) + \w_{out}(s)} + \\+ \frac{ C_1^*(p) C_1^*(q) C_1^*(r) C_1(s) C_2^*(p) C_1(q) C_1(r) C_1^*(s) - C_2^*(p) C_2^*(q) C_2^*(r) C_2(s) C_1^*(p) C_2(q) C_2(r) C_2^*(s)  }{\w_{out}(p) + \w_{out}(q) + \w_{out}(r) - \w_{out}(s)} + \\ + 2\frac{C_1^*(p) C_2^*(q) C_1^*(r) C_2(s) C_2^*(p) C_2(q) C_1(r) C_2^*(s) - C_2^*(p) C_1^*(q) C_2^*(r) C_1(s) C_1^*(p) C_1(q) C_2(r) C_1^*(s) }{\w_{out}(p) - \w_{out}(q) + \w_{out}(r) + \w_{out}(s)} + \\+ 2\frac{ C_1^*(p) C_1^*(q) C_2^*(r) C_1(s) C_2^*(p) C_1(q) C_2(r) C_1^*(s) - C_2^*(p) C_2^*(q) C_1^*(r) C_2(s) C_1^*(p) C_2(q) C_1(r) C_2^*(s)  }{\w_{out}(p) + \w_{out}(q) - \w_{out}(r) - \w_{out}(s)}  + \\+ \frac{ C_1^*(p) C_2^*(q) C_2^*(r) C_2(s) C_2^*(p) C_2(q) C_2(r) C_2^*(s) - C_2^*(p) C_1^*(q) C_1^*(r) C_1(s) C_1^*(p) C_1(q) C_1(r) C_1^*(s)  }{\w_{out}(p) - \w_{out}(q) - \w_{out}(r) + \w_{out}(s)} + \\+ \frac{ C_1^*(p) C_2^*(q) C_2^*(r) C_1(s) C_2^*(p) C_2(q) C_2(r) C_1^*(s) - C_2^*(p) C_1^*(q) C_1^*(r) C_2(s) C_1^*(p) C_1(q) C_1(r) C_2^*(s) }{\w_{out}(p) - \w_{out}(q) - \w_{out}(r) - \w_{out}(s)} \bigg).
\end{multline*}

In all, due to the space-time expansion, the in-modes (that are single waves in the remote past) behave as linear superpositions of plane waves in the remote future. It is the interference between these plane waves at the future infinity which causes the appearance of the growing with time contributions in the occupation number and anomalous average. This growth is signaling in a change of the initial state of the interacting theory. As we see, in the present case that is caused by the expansion of the space-time. But here we see only the leading effect in the second loop order.  

Namely we see that even if $\lambda$ is very small, loop corrections to the occupation number and anomalous average (to the two-point functions) is becoming strong after a long enough evolution time. This means the breakdown of the perturbation theory. Then to understand the physics one has at least to resum the leading growing contributions from all loops. That is the problem for the future work. It is a much harder problem than the resummation of the bubble diagrams because it involves kinetic processes both in occupation number and anomalous average.

Let us stress here the secular growth in $n_p$ and $\kappa_p$ is not solely due to the expansion of the geometry. In fact, for the same scalar theory in the background of the usual $(d+1)$-dimensional flat spacetime one encounters a similar phenomenon. If the initial state is not the Fock space ground state, but rather contains some non-zero occupation number $n_p$, which is different from the Planckian spectrum, then the two-loop correction to the Keldysh propagator yields the following contribution to the occupation number:
\begin{multline} \label{mink}
    n_p^{(2)} \sim \lambda^2 (t-t_0) \int d^dq \,d^dr\, d^ds\, \delta(\w_p+\w_q-\w_r-\w_s) \times \\ \times \bigg[ (1+n_p)(1+n_q)n_r n_s - n_p n_q (1+n_r) (1+n_s)  \bigg],
\end{multline}
where $\w_p = \sqrt{\vec{p}^2+m^2}$ and $t_0$ is the position of the initial Cauchy surface. See, e.g.,  \cite{Akhmedov:2021rhq} and references therein.

Note that the r.h.s of this relation vanishes only for the Planckian distribution $n_p = (\exp(\beta \w_p)-1)^{-1}$, which includes the Fock space ground state $n_p = 0$.  Otherwise we encounter simultaneously the secular growth and secular divergence. The latter is the dependence on $t_0$, which cannot be taken to the past infinity, because otherwise the loop correction will be infinite. Unlike the stationary situation, now we find the explicit dependence on the initial Cauchy surface. Thus, in the situation under consideration the secular growth in the occupation number comes from the fact that due to the selfinteraction, $\lambda \, \phi^4$, the occupation numbers start to change in time --- we see their redistribution due to particle scattering processes. 

Somewhat similar phenomenon we encounter in the expanding universe. In fact, the initial state that we take is not a ground state of the time dependent Hamiltonian of the theory that we consider. Furthermore, due to the time dependence of the background we also encounter the generation of the anomalous averages, even if they have been zero initially. And we of course see many more different kinetic processes, which are forbidden in the flat space-time due to the energy conservation. 

\subsection{A comment about out-modes}

Now let us consider the loop corrections for the out-modes (for the initial state constructed over the Fock space out-state). In fact, the space-time expansion could have started from any quantum state rather than from the thermal in-state.

Performing similar operations as for the in-modes, one obtains that in the limit \eqref{limit} the loop corrections to the occupation number and anomalous quantum average for the out-modes are given by the same expressions \eqref{n_phi4_modes} and \eqref{kappa_4_modes}, correspondingly, but with the in-modes $g_p^{in}(\eta)$ replaced by the out-modes $g_p^{out}(\eta)$.

To continue we again divide the region of integration over $\eta_3$ and $\eta_4$ into three time intervals and consider the contribution only from the distant future region, which potentially can provide the leading contribution in the limit $\eta \to + \infty$, if $\eta_0$ is kept finite. We will see in a moment that it is very important to keep $\eta_0$ finite for out-states.

Plugging asymptotic form of the out-modes \eqref{tad_out_asympt} in the future region into the expression for the occupation number, one obtains that
\begin{multline}
    \bar{n}_p^{(2)} \approx\frac{\lambda^2 (A+B)^{2} \, \eta}{ 64 \,\pi} \int \frac{dq \, dr \, ds \, \delta\Big({p}-{q}-{r}-{s}\Big)}{\w_{out}(p)\w_{out}(q)\w_{out}(r) \w_{out}(s)} \delta\Big(\w_{out}(p) + \w_{out}(q) - \w_{out}(r) - \w_{out}(s)\Big) \times \\ \times \Big[ (1+\bar{n}_p)(1+\bar{n}_{q})\bar{n}_r \bar{n}_s - \bar{n}_p \bar{n}_{q} (1+\bar{n}_r)(1+\bar{n}_s) \Big].
\end{multline}
However, as it was pointed out in \eqref{mom_sol1} and \eqref{mom_sol2}, due to the presence of the delta-functions $\delta(p-q-r-s)$ and 
$\delta(\w_{out}(p) + \w_{out}(q) - \w_{out}(r) - \w_{out}(s))$ the expression 
$\Big[ (1+\bar{n}_p)(1+\bar{n}_{q})\bar{n}_r \bar{n}_s - \bar{n}_p \bar{n}_{q} (1+\bar{n}_r)(1+\bar{n}_s) \Big]$ 
is equal to zero. Therefore, the loop corrections to the occupation number for the out-modes are not growing as $\eta \to + \infty$. For the same reason we have neglected contribution from the remote past region for the in-modes, because the latter contribution was not growing as $\eta_0 \to - \infty$.
Furthermore, plugging \eqref{tad_out_asympt} into the expression for the anomalous quantum average, one also obtains that the loop corrections to the anomalous quantum average for the out-modes does not grow in the limit \eqref{limit}. 

However, let us stress here that for the out-modes one obtains the infrared catastrophe. In fact, if $\eta_0 \to -\infty$, then the contribution from the remote past region, $(\eta_0,\Bar{\eta})$, is growing as $\eta_0 \to - \infty$. That happens for the same reason as why there is the secular growth for the in-modes in the limit $\eta \to +\infty$. Due to this divergence the initial Cauchy surface, $\eta_0$, cannot be taken to the past infinity. Otherwise the loop correction to the propagator will be infinite even after the implementation of the UV cutoff. Similar situation one encounters in global de Sitter space-time \cite{Akhmedov:2021rhq} (see also \cite{Krotov:2010ma}, \cite{Akhmedov:2013vka}). 

But let us stress that if $\eta_0$ is taken long before the beginning of expansion we essentially encounter a situation with a coherent initial state in flat space-time. Namely, we encounter a situation with such a state that leads to a non-zero initial anomalous average. This situation was considered in \cite{Akhmedov:2021vfs}. If $\eta_0\to - \infty$, one can expect that the system will get thermalized before the beginning of the expansion. 

\section{Conclusions and discussion} \label{set}

So far we have essentially considered corrections to the two-point functions: as one can see from (\ref{gk_modes_in}) the Keldysh propagator encodes the occupation numbers and the anomalous averages. We have seen that in $\lambda \phi^4$ theory the occupation numbers and anomalous averages for the in-modes grow with the increase of the average time of the two-point function. To see what physical consequences this effect leads to in this section we calculate the corrections to the expectation value of the stress energy tensor. 

To calculate the expectation value we use the standard technique of \cite{Bernard:1977pq}. For simplicity we assume that the initial occupation number is zero, as in classic texts. But we have to take into account that now due to the self-interactions the state in the asymptotic region $\eta \to \infty$ acquires non-zero occupation number and anomalous quantum average. We also need to take into account the mass renormalization coming from the summation of the leading tadpole diagrams.

We use the Pauli-Villars regularization scheme, i.e. we introduce the regulator fields. Then, after the resummation of the tadpole diagrams, one obtains the following effective Lagrangian:
\begin{multline} \label{lagr_tot}
    \mathcal{L}_{\text{eff}} = \frac{1}{2} \, g^{\mu \nu} \bigg( \partial_{\mu} \phi \, \partial_{\nu} \phi + \partial_{\mu} \psi \, \partial_{\nu} \psi + 2 \, \partial_{\mu} \chi^{\dagger} \, \partial_{\nu} \chi \bigg) - \\-\frac{1}{2} \bigg( \Tilde{m}^2 \phi^2 + (2M^2-\Tilde{m}^2)\, \psi^2 + 2M^2 \chi^{\dagger} \chi \bigg) - \frac{\lambda}{4!} \phi^4 + Z(\Tilde{m}^2,M^2),
\end{multline}
where $\Tilde{m}$ is the renormalized mass (defined in \eqref{mass_ren}), $\psi$ is a commuting field, while $\chi$ and $\chi^{\dagger}$ are anticommuting fields, and $Z(\Tilde{m}^2,M^2)$ is a counterterm (cosmological constant renormalization). We do not take into account here the $\lambda \phi^4$ term, because its contribution does not affect our conclusions substantially.  

Varying the action with the Lagrangian \eqref{lagr_tot}  with respect to the metric tensor, one obtains that the stress-energy tensor is given by
\begin{multline}
    T^{\mu \nu} \equiv -\frac{2}{\sqrt{-g}} \frac{\delta S}{\delta g_{\mu \nu}} = -\frac{1}{2} \, \bigg(g^{\mu \nu} g^{\rho \sigma} - g^{\mu \rho} g^{\nu \sigma} - g^{\nu \rho} g^{\mu \sigma}  \bigg) \bigg( \partial_{\rho} \phi \, \partial_{\sigma} \phi + \partial_{\rho} \psi \, \partial_{\sigma} \psi + 2 \, \partial_{\rho} \chi^{\dagger} \, \partial_{\sigma} \chi \bigg) + \\+\frac{1}{2} g^{\mu \nu} \bigg( \Tilde{m}^2 \phi^2 + (2M^2-\Tilde{m}^2)\, \psi^2 + 2M^2 \chi^{\dagger} \chi \bigg) - \frac{\lambda}{4!} g^{\mu \nu} \phi^4 - g^{\mu \nu} Z(\Tilde{m}^2,M^2) .
\end{multline}
Now we consider the time evolution of the Fock space ground state for the in-modes from $\eta \to -\infty$ to $\eta \to +\infty$. We assume that the interaction term is turned on adiabatically, so that in the distant past region we have the free theory. Then the role of the interaction term is to create non-zero occupation number and anomalous quantum average during the expansion of the space-time. 

Because at past infinity we start with the empty space, we can fix the counterterm $Z(\Tilde{m}^2,M^2)$ \textcolor{black}{by requiring that the stress-energy tensor at past infinity, $\eta \to -\infty$, is zero}. Evaluating the expectation value of the stress energy tensor at past infinity $\eta \to -\infty$ for such an in-state $|\text{in}\rangle$ that
\begin{equation}
    \langle \text{in}| a^{\dagger}_q a_k |\text{in}\rangle =0, \qquad \langle \text{in} | a_q a_k| \text{in} \rangle =0, \quad {\rm as} \quad \eta \to -\infty,
\end{equation}
one finds that the counterterm should have the following form (see \cite{Birrell:1982ix}, \cite{Bernard:1977pq} for the details of the calculation):
\begin{equation}
    Z(\Tilde{m}^2,M^2) = - \frac{1}{8\pi} \bigg( \Tilde{m}^2 \log \Tilde{m}^2 + (2M^2-\Tilde{m}^2) \log (2M^2-\Tilde{m}^2) -2M^2 \log M^2 \bigg).
\end{equation}
However, as it was shown in the previous sections, the same in-ground-state $| \text{in} \rangle$ after the evolution transforms into a state with a non-zero occupation number and anomalous quantum average in the future infinity, $\eta \to +\infty$, i.e.
\begin{equation}
    \langle \text{in}| S^+ \, a^{\dagger}_q a_k \, S |\text{in}\rangle_{loop} =n_k^{(2)} \delta(q-k), \qquad \langle \text{in} | S^+ \, a_q a_k \, S| \text{in} \rangle_{loop} = \kappa_k^{(2)} \delta(k+q), \quad {\rm as} \quad \eta \to +\infty,
\end{equation}
where the explicit expressions for $n_k^{(2)}$ and $\kappa_k^{(2)}$ can be derived from \eqref{n_p_fin} and \eqref{kappa_4_fin} by putting initial occupation number to be zero, i.e. $n_p=0$, as we start here from the Fock space ground state for the in-modes. Note that in this paper we have expanded the evolution operator $S$ to the second loop order only.

Then, computing the expectation value of the stress-energy tensor in the distant future and taking the regulator field masses to infinity ($M \to \infty$), one obtains that:
\begin{align}
    &\langle  \text{in} | T^{\mu \nu}|  \text{in} \rangle = \langle  \text{in}| T^{\mu \nu}_0|  \text{in} \rangle + \langle  \text{in}| T^{\mu \nu}_{\text{loop}}| \text{in} \rangle,\nonumber \\
    &\langle  \text{in} | T^{\mu \nu}_0 |  \text{in} \rangle \xrightarrow[\eta \to +\infty]{} \int^{\infty}_{-\infty} \frac{dk}{4\pi \Tilde{\w}_{out}} k^{\mu} k^{\nu} \bigg( \frac{\Tilde{\w}_{out}}{\Tilde{\w}_{in}} \big( |C_1|^2 + |C_2|^2 \big) -1 \bigg) ,\nonumber \\
    &\langle  \text{in} | T^{\mu \nu}_{\text{loop}}|  \text{in} \rangle \xrightarrow[\eta \to +\infty]{} \int^{\infty}_{-\infty} \frac{dk}{2\pi \Tilde{\w}_{in}} k^{\mu} k^{\nu} \bigg[ n_k^{(2)} \big( |C_1|^2 + |C_2|^2 \big) + \kappa_k^{(2)} C_1 C_2 + \kappa_k^{(2)*} C_1^* C_2^* \bigg],
\end{align}
where
\begin{equation}
    k^{\mu} = \bigg( \frac{\Tilde{\w}_{out}}{A+B}, \frac{k}{A+B} \bigg).
\end{equation}
Here the term $\langle  \text{in} | T^{\mu \nu}_0 |  \text{in} \rangle$ is the contribution of the gaussian theory (given by the Lagrangian \eqref{lagr_tot} with self-interaction switched off), while the term $\langle  \text{in} | T^{\mu \nu}_{\text{loop}}|  \text{in} \rangle$ arises from the loop corrections to the occupation number and anomalous quantum average, and $\Tilde{\w}_{in}$ and $\Tilde{\w}_{out}$ are defined in \eqref{tilde_omega}. While $\langle  \text{in} | T^{\mu \nu}_0 |  \text{in} \rangle$ is due to amplification of the zero-point fluctuations, $\langle  \text{in} | T^{\mu \nu}_{\text{loop}}|  \text{in} \rangle$ is due to the excitation of the higher levels --- due to the change of the state of the theory.
Note, at the same time, that the term $\langle  \text{in} | T^{\mu \nu}_{\text{loop}}|  \text{in} \rangle $ is proportional to $\lambda^2 \, \eta$. Hence, even for small $\lambda$ this term is comparable to the tree level result $\langle  \text{in} | T^{\mu \nu}_0 |  \text{in} \rangle$ as $\eta \to \infty$. 

Thus, the well known textbook tree-level contribution to the expectation value of the stress energy tensor is strongly corrected by the quantum loop effects in self-interacting theories. That signals a violation of the perturbation theory expansion, because higher loops deliver corrections of the same order $\left(\lambda^2 \, \eta\right)^n \sim 1$. Hence, to understand the physical consequences of the effects that we consider here one has to resum at least the leading contributions from all loops. Let us stress that the result of the resummation may depend on the initial conditions. See, e.g. \cite{Akhmedov:2020haq, Akhmedov:2019ubc, Akhmedov:2017ooy, Akhmedov:2013xka, Trunin:2021lwg} for a similar resummation which was performed in analogous situations.  We will do the loop resummation in the situation under consideration in the future work.

\section{Acknowledgements}
We would like to thank A. Alexandrov, D. Diakonov, K. Gubarev, K. Kazarnovskii, and D. Trunin for useful discussions. This work was supported by the Foundation for the Advancement of Theoretical Physics and Mathematics “BASIS” grant, by RFBR grants 19-02-00815 and 21-52-52004, and by Russian Ministry of education and science.

\begin{appendices}
\numberwithin{equation}{section}

\renewcommand\theequation{A.\arabic{equation}}
\section{Generalization to higher dimensions}

In this Appendix we briefly discuss the situation in higher dimensional space-times. We want to show that the phenomenon that we are discussing here is not specific only to the two dimensional case. Namely, we consider $(d+1)$-dimensional space-time with the metric:
\begin{equation}
    ds^2 = C(\eta) (d\eta^2 - d \Vec{x}^2), \qquad \Vec{x} = (x_1,...,x_d),
\end{equation}
with the same conformal factor $C(\eta) = A + B\tanh(\rho \eta)$ as in two-dimensional case.
Separating the variables as
\begin{equation}
    u_k(\eta,\Vec{x}) = e^{i\Vec{k}\Vec{x}} g_k(\eta),
\end{equation}
and plugging these mode functions into the equation of motion, for the temporal part of the modes one obtains the equation as follows
\begin{equation}
    \frac{d^2 g_k(\eta)}{d\eta^2} + \frac{d-1}{2} \frac{C'(\eta)}{C(\eta)} \frac{\partial g_k(\eta)}{\partial \eta} + \Big[k^2+m^2 C(\eta) \Big] g_k(\eta)=0.
\end{equation}
Here one can get rid of the first order derivative and rewrite this equation in the form of the oscillator with a time-dependent frequency:
\begin{equation}
    \bigg[ \frac{d^2}{d\eta^2} -\frac{(d-1)(d-5)}{16} \frac{(C'(\eta))^2}{C^2(\eta)} - \frac{d-1}{4} \frac{C''(\eta)}{C(\eta)} + k^2 + m^2 C(\eta) \bigg] \, g_k(\eta) \, C^{\frac{d-1}{4}} (\eta)  = 0.
\end{equation}
Exact solutions of this equation are given by the Heun functions. Their explicit from is not necessary for our purposes. In fact, we may distinguish solution, which behaves in the asymptotic regions as:
\begin{align} \label{in_modes_asympt_hd}
    u_k^{in}(\eta,\Vec{x}) \sim 
    \begin{cases}
        \frac{1}{\sqrt{4\pi \w_{in}}} {C(\eta)^{-\frac{d-1}{4}}} e^{i\Vec{k} \Vec{x}-i\w_{in}\eta}, \qquad \eta \to -\infty \\
        \frac{1}{\sqrt{4\pi \w_{in}}} {C(\eta)^{-\frac{d-1}{4}}} e^{i\Vec{k} \Vec{x}} \bigg[ \Tilde{C}_1 e^{-i\w_{out}\eta} + \Tilde{C}_2 e^{i\w_{out}\eta} \bigg], \qquad \eta \to +\infty,
    \end{cases}
\end{align}
with some coefficients $\Tilde{C}_1$ and $\Tilde{C}_2$.
The mode decomposition of the field operator as usual has the form:
\begin{equation}
    \phi(\eta,\Vec{x}) = \int^{\infty}_{-\infty} d^d k \Big[ a_k u_k^{in}(\eta,\vec{x}) + a_k^{\dagger} u_k^{in*}(\eta,\vec{x}) \Big].
\end{equation}
As in the two-dimensional case, the sum of the leading tadpole diagrams contributes to the mass and mode renormalization. Here we imply that the mass is already renormalized and we aim to demonstrate that the loop corrections to occupation number and anomalous quantum average contain growing with time contributions.

From here it is tedious but straightforward to deduce that the expressions for the occupation number and anomalous quantum averages are essentially the same as in the two-dimensional case, but with the different numerical factors and small modifications. 
%For the qubic interaction the two-loop contribution in the limit (\ref{limit}) is still zero due to the fact that the system of equations
%\begin{align}
%    \begin{cases}
%        \vec{p}=\vec{q}+\vec{r},\\
%        \pm \w_{out}(p) \pm \w_{out}(q) \pm \w_{out}(r)=0.
%    \end{cases}
%\end{align}
%still does not have a real solutions (one may check it by going to the rest frame of one of the ``particles''). 

E.g. for the occupation number one obtains the following expression
\begin{multline}
    n_p^{(2)} \approx (2\pi)^{2d} \frac{\lambda^2}{6} \int^{\eta}_{\eta_0}d\eta_3 C(\eta_3)^{\frac{d+1}{2}} \int^{\eta}_{\eta_0}d\eta_4 C(\eta_4)^{\frac{d+1}{2}} \int d^dq\,d^dr\,d^ds\, \delta\Big(\vec{p}-\vec{q}-\vec{r}-\vec{s}\Big) g_p^{in*}(\eta_3) g_p^{in}(\eta_4) \times \\ \times
    \bigg[   g_q^{in*}(\eta_3)g_q^{in}(\eta_4) g_r^{in*}(\eta_3)g_r^{in}(\eta_4) g_s^{in*}(\eta_3)g_s^{in}(\eta_4) \bigg( (1+n_p)(1+n_{q})(1+n_{r})(1+n_{s}) - n_p n_{q} n_{r} n_{s} \bigg) +\\
    + 3g_q^{in*}(\eta_3)g_q^{in}(\eta_4) g_r^{in*}(\eta_3)g_r^{in}(\eta_4) g_s^{in}(\eta_3)g_s^{in*}(\eta_4) \bigg( (1+n_p)(1+n_{q})(1+n_{r})n_s - n_p n_{q} n_{r} (1+n_s) \bigg)+\\
    +3g_q^{in*}(\eta_3)g_q^{in}(\eta_4) g_r^{in}(\eta_3)g_r^{in*}(\eta_4) g_s^{in}(\eta_3)g_s^{in*}(\eta_4) \bigg( (1+n_p)(1+n_{q})n_r n_s - n_p n_{q} (1+n_r) (1+n_s) \bigg)+\\
    + g_q^{in}(\eta_3)g_q^{in*}(\eta_4) g_r^{in}(\eta_3)g_r^{in*}(\eta_4) g_s^{in}(\eta_3)g_s^{in*}(\eta_4) \bigg( (1+n_p)n_q n_r n_s - n_p (1+n_q)(1+n_r)(1+n_s) \bigg)
    \bigg].
\end{multline}
Evaluating integrals over $\eta_3$ and $\eta_4$ as in the two-dimensional case, it is straightforward to find that:
\begin{align}
    n_p^{(2)} \approx \frac{\lambda^2 (A+B)^{3-d} \, \eta}{ 2^{8-2d} \,\pi^{3-2d}} \int \frac{d^dq \, d^dr \, d^ds \, \delta\Big(\vec{p}-\vec{q}-\vec{r}-\vec{s}\Big)}{\w_{in}(p)\w_{in}(q)\w_{in}(r) \w_{in}(s)} \delta(\w_{out}(p) + \w_{out}(q) - \w_{out}(r) - \w_{out}(s)) \times \nonumber
     \\ \times \bigg\{  \Tilde{\mathcal{N}}_1(p,q,r,s)
    \Big[ (1+n_p)(1+n_{q})(1+n_{r})(1+n_{s}) - n_p n_{q} n_{r} n_{s} \Big] +\nonumber \\+\Tilde{\mathcal{N}}_2(p,q,r,s) \Big[ (1+n_p)(1+n_{q})(1+n_{r})n_s - n_p n_{q} n_{r} (1+n_s) \Big] + \nonumber \\+ \Tilde{\mathcal{N}}_3(p,q,r,s) \Big[ (1+n_p)n_q(1+n_{r})(1+n_{s}) - n_p (1+n_q) n_{r}  n_{s}  \Big] +\nonumber \\+ \Tilde{\mathcal{N}}_4(p,q,r,s) \Big[ (1+n_p)(1+n_{q})n_r n_s - n_p n_{q} (1+n_r)(1+n_s) \Big] +\nonumber \\+\Tilde{\mathcal{N}}_5(p,q,r,s) \Big[ (1+n_p)n_q n_r (1+n_{s})  - n_p (1+n_q) (1+n_r) n_{s}\Big] +\nonumber \\+ \Tilde{\mathcal{N}}_6(p,q,r,s) \Big[ (1+n_p)n_q n_r n_s - n_p (1+n_q)(1+n_r)(1+n_s) \Big] \bigg\},
\end{align}
where the quantities $\Tilde{\mathcal{N}}_{1,2,3,4,5,6}(p,q,r,s)$ are given by
\begin{align*}
    &\Tilde{\mathcal{N}}_1(p,q,r,s) = | \Tilde{C}_1 (p) \Tilde{C}_1(q)\Tilde{C}_2(r) \Tilde{C}_2(s)|^2+|\Tilde{C}_2(p) \Tilde{C}_2(q) \Tilde{C}_1(r) \Tilde{C}_1(s) |^2,\\
    &\Tilde{\mathcal{N}}_2(p,q,r,s) = 2 \bigg( |\Tilde{C}_1(p) \Tilde{C}_1(q) \Tilde{C}_2(r) \Tilde{C}_1(s)|^2 +|\Tilde{C}_2(p) \Tilde{C}_2(q) \Tilde{C}_1(r) \Tilde{C}_2(s) |^2 \bigg),\\
    &\Tilde{\mathcal{N}}_3(p,q,r,s) = | \Tilde{C}_1(p)  \Tilde{C}_2(q) \Tilde{C}_2(r) \Tilde{C}_2(s)|^2  +|\Tilde{C}_2(p) \Tilde{C}_1(q) \Tilde{C}_1(r) \Tilde{C}_1(s)|^2,\\
    &\Tilde{\mathcal{N}}_4(p,q,r,s) = |\Tilde{C}_1(p) \Tilde{C}_1(q) \Tilde{C}_1(r) \Tilde{C}_1(s)|^2  +|\Tilde{C}_2(p) \Tilde{C}_2(q) \Tilde{C}_2(r) \Tilde{C}_2(s) |^2,\\
    &\Tilde{\mathcal{N}}_5(p,q,r,s) = 2 \bigg( |\Tilde{C}_1(p) \Tilde{C}_2(q) \Tilde{C}_1(r) \Tilde{C}_2(s)|^2 +|\Tilde{C}_2(p) \Tilde{C}_1(q) \Tilde{C}_2(r) \Tilde{C}_1(s)|^2 \bigg),\\
    &\Tilde{\mathcal{N}}_6(p,q,r,s) = |\Tilde{C}_1(p) \Tilde{C}_2(q) \Tilde{C}_1(r) \Tilde{C}_1(s)|^2 +|\Tilde{C}_2(p) \Tilde{C}_1(q) \Tilde{C}_2(r) \Tilde{C}_2(s) |^2.
\end{align*}

One should note here that the only difference in higher dimensions, as compared to the two-dimensional case, is that the internal momenta now are $d$-dimensional, and also the numerical prefactor of the r.h.s. is different. However, the character of the secular growth is linear in any number of dimensions, and the form of the ``collision integral'' is also the same. The same fact holds for the anomalous quantum average --- the expression for it is similar to \eqref{kappa_4_fin}. 
%\eqref{I_1}, \eqref{I_2}.

At the same time, for the out-modes the correction coming from the future region is as follows:
\begin{multline}
    \bar{n}_p^{(2)} \approx\frac{\lambda^2 (A+B)^{3-d} \, \eta}{ 2^{8-2d} \,\pi^{3-2d}} \int \frac{d^dq \, d^dr \, d^ds \, \delta\Big(\vec{p}-\vec{q}-\vec{r}-\vec{s}\Big)}{\w_{out}(p)\w_{out}(q)\w_{out}(r) \w_{out}(s)} \delta\Big(\w_{out}(p) + \w_{out}(q) - \w_{out}(r) - \w_{out}(s)\Big) \times \\ \times \Big[ (1+\bar{n}_p)(1+\bar{n}_{q})\bar{n}_r \bar{n}_s - \bar{n}_p \bar{n}_{q} (1+\bar{n}_r)(1+\bar{n}_s) \Big].
\end{multline}
Here we keep $\eta_0$ finite and neglect the contribution from the past region. Otherwise, if $\eta_0 \to - \infty$ we obtain the infrared catastrophe, as was discussed above.

In all, the situation in higher dimensions is not conceptually different from the two-dimensional case, if one considers first few corrections to the stress energy tensor in the massive scalar theory.

\end{appendices}

\end{document}